\documentclass[12pt,onecolumn]{IEEEtran}
\usepackage{amsmath,amsfonts,amssymb,caption,graphicx,dsfont,color}
\input{epsf}

\def\ngi{{n \rightarrow \infty}}
\def\VAR {{\mathsf{Var}}}

\newcommand{\set}[1]{\mathcal{#1}}   
\newcommand{\Reals}{\mathbb R}  

\newtheorem{definition}{Definition}
\newtheorem{theorem}{Theorem}

\newtheorem{lemma}{Lemma}

\newtheorem{property}{Property}
\newtheorem{example}{Example}

\title{
Lossless Data Compression\\ at Finite Blocklengths}

\author{\IEEEauthorblockN{Ioannis Kontoyiannis}\\
\IEEEauthorblockA{{\small Department of Informatics}\\
{\small Athens University of Economics and Business}\\
{\small Athens 10434, Greece}}\\
{\small\texttt{yiannis@aueb.gr}}\\
\medskip
\IEEEauthorblockN{Sergio Verd\'{u}}\\
\IEEEauthorblockA{{\small Department of Electrical Engineering}\\
{\small Princeton University}\\
{\small Princeton, New Jersey 08544, USA}\\
{\small\texttt{verdu@princeton.edu}}\\
}
}

\begin{document}
\date{\today}

\maketitle
\begin{abstract}
This paper provides an extensive 
study of the behavior of the 
best achievable rate (and other related
fundamental limits) in variable-length lossless compression.
In the non-asymptotic regime,
the fundamental limits of fixed-to-variable 
lossless compression with and without prefix 
constraints are shown to be tightly coupled.
Several precise, quantitative bounds are derived, 
connecting the distribution of the optimal codelengths
to the source information spectrum, and an
exact analysis of the best achievable rate
for arbitrary sources is given.

Fine asymptotic results are proved for 
arbitrary (not necessarily prefix) compressors
on general mixing sources.
Non-asymptotic, explicit Gaussian approximation 
bounds are established for the best achievable rate
on Markov sources.
The source dispersion and the source varentropy
rate are defined and characterized.
Together with the entropy rate, the varentropy rate 
serves to tightly approximate the fundamental 
non-asymptotic limits of fixed-to-variable 
compression for all but very small blocklengths.
\end{abstract}


\newpage

\section{Fundamental Limits}\label{sec:1}
\subsection{The optimum fixed-to-variable code}\label{sec:ofv}

A fixed-to-variable compressor for a finite alphabet $\set{A}$ is an 
injective function,
\begin{eqnarray}\label{strings}
{\mathsf f}_n \colon {\cal A}^n \to \{ 0 , 1 \}^* = 
\{ \varnothing, 0, 1, 00, 01, 10, 11, 000, 001, \ldots\} .
\end{eqnarray}
The length of a string $a \in \{0,1\}^*$ is denoted by $\ell (a) $.
Therefore, a block (or file) of $n$ symbols 
$a^n=(a_1,a_2,\ldots,a_n) \in \set{A}^n$ is losslessly 
compressed by ${\mathsf f}_n$ into a binary string 
whose length is 
$\ell ( {\mathsf f}_n (a^n) )$ bits.

When the file 
$X^n=(X_1,X_2,\ldots,X_n)$ to be compressed 
is generated by a probability law  $P_{X^n}$,
a basic information-theoretic object of study
is the distribution of the rate of the optimal 
compressor, seen as a function of the
blocklength $n$ and the distribution $P_{X^n}$.
The best achievable compression
performance at 
finite blocklengths is characterized 
by fundamental limits, including:
\begin{enumerate}
\item $R^* (n , \epsilon )$: The lowest rate
$R$ such that the compression rate 
of the best code exceeds $R$ with probability 
not greater than $\epsilon$:
\begin{eqnarray} \label{fl1}
\min_{{\mathsf f}_n} \mathbb{P} [ \ell ( {\mathsf f}_n (X^n) ) >  n  R  ] \leq \epsilon.
\end{eqnarray}
\item  
$\epsilon^*(n,k)$: The smallest possible
excess-rate probability, namely, the
probability that the compressed length is  
greater than or equal to $k$:
\begin{eqnarray}\label{fl2}
\epsilon^* ( n , k ) = \min_{{\mathsf f}_n} \mathbb{P}
[ \ell ( {\mathsf f}_n (X^n) ) \geq  k  ] .
\end{eqnarray}
\item $n^*( R , \epsilon )$: The smallest blocklength
at which compression at rate $R$ is possible with
probability at least $1-\epsilon$; in other words,
the minimum $n$ required for \eqref{fl1} to hold.
\item $\bar{R}(n)$: The minimal average compression rate:
\begin{eqnarray}\label{fl3}
\bar{R} (n ) &=&  
\frac{1}{n} \min_{{\mathsf f}_n} \mathbb{E} 
	[ \ell ( {\mathsf f}_n (X^n) )] \\
&=& \frac{1}{n} \sum_{k=1}^\infty \epsilon^* (n, k).  \label{sne}
\end{eqnarray}
\end{enumerate}

Naturally, the 
fundamental limits in 1),  2) and 3) are equivalent 
in the sense that knowledge of one of them 
(as a function of its parameters) determines the other two. 
For example, 
\begin{eqnarray} \label{renk}
R^* (n, \epsilon ) = \frac{k}{n} ~~\mbox{if and only if}~ \epsilon^* (n, k) \leq \epsilon < \epsilon^* (n, k-1).
\end{eqnarray}
As for 4), we observe that, together 
with \eqref{sne} and the fact that
$\epsilon^* (n, 0) = 1$, \eqref{renk} results in:
\begin{eqnarray}
\bar{R} (n ) = \int_0^1 R^* (n , x ) \, dx  -\frac{1}{n}.  \label{90d}
\end{eqnarray}

The minima in the fundamental limits \eqref{fl1}, \eqref{fl2}, \eqref{fl3} 
are achieved by 
an optimal compressor ${\mathsf f}_n^*$ that assigns the elements 
of ${\cal A}^n$ ordered in decreasing probabilities
to the elements in $\{0,1\}^*$ ordered lexicographically as in \eqref {strings}.
In particular,
\begin{equation}
\bar{R} (n ) 
= \frac{1}{n} \mathbb{E} 
	[ \ell ( {\mathsf f}^*_n (X^n) )],
\label{eq:Rbarn}
\end{equation}
and,
\begin{equation}
R^*(n,\epsilon)
\;\mbox{is the smallest $R$ s.t.}\;
\mathbb{P} 
[ \ell ( {\mathsf f}^*_n (X^n) ) >  n  R  ] \leq \epsilon,
\label{eq:optimalf}
\end{equation}
where the optimal
compressor ${\mathsf f}_n^*$ is described precisely as:

\medskip

\begin{property}\label{property:1}
For every $k = 1, \ldots , \lfloor \log_2 ( 1 + | \mathcal{A} |^n ) \rfloor  $, 
any optimal code ${\mathsf f}_n^*$ assigns strings of length $0, 1, 2, \ldots , k-1$ to
each of the 
\begin{eqnarray}
1 + 2 + 4 + \ldots + 2^{k-1} = 2^k - 1,
\end{eqnarray}
most likely  elements
of ${\cal A}^n$.  
If  $\log_2 ( 1 +  | \mathcal{A} |^n ) $ 
is not an integer, then ${\mathsf  f}_n^*$ assigns
strings of length $ \lfloor \log_2 ( 1 + | \mathcal{A} |^n ) \rfloor $ 
to the least likely $ | \set{A} |^n + 1 - 2^{\lfloor \log_2 
( 1 + | \set{A} |^n ) \rfloor} $
 elements in  $\set{A}^n$.
\end{property}

\medskip

Note that 
Property \ref{property:1} is a necessary and sufficient condition for optimality, which does not determine
${\mathsf  f}_n^*$ uniquely: not only does it not specify how to break ties among probabilities but any swap 
between two codewords of the same length preserves optimality. As in the following example, it is convenient, however, to think of ${\mathsf  f}_n^*$
as the unique compressor constructed by breaking ties lexicographically and by assigning the elements of  $\{ 0 , 1 \}^*$
in the lexicographic order of \eqref{strings}. 

\medskip

\begin{example}
Suppose $n=4$, 
$\set{A} = \{  \circ, \bullet \}  $, 
and the source is memoryless 
with 
$\mathbb{P} [ X = \bullet ] > \mathbb{P} [ X = \circ ] $. 
Then the following compressor is optimal:
\begin{eqnarray}
{\mathsf  f}_4^*(\bullet\,\bullet\,\bullet\,\,\bullet )& = &\varnothing \nonumber\\
{\mathsf  f}_4^*(\bullet\,\bullet\,\bullet\,\, \circ) &=& \mathtt{ 0 }\nonumber \\
{\mathsf  f}_4^*(\bullet\,\bullet\,\circ\,\,\bullet) &=& \mathtt{1 }\nonumber \\
{\mathsf  f}_4^*(\bullet\,\circ\,\bullet\,\,\bullet) &=& \mathtt{00 }\nonumber \\
{\mathsf  f}_4^*(\circ\,\bullet\,\bullet\,\,\bullet) &=& \mathtt{01 }\nonumber \\
{\mathsf  f}_4^*(\circ\,\circ\,\bullet\,\,\bullet) &=& \mathtt{10 } \nonumber \\
{\mathsf  f}_4^*(\circ\,\bullet\,\bullet\,\,\circ) &=& \mathtt{11 }\nonumber \\
{\mathsf  f}_4^*(\circ\,\bullet\,\circ\,\,\bullet) &=& \mathtt{000 } \nonumber \\
{\mathsf  f}_4^*(\bullet\,\bullet\,\circ\,\,\circ) &=& \mathtt{001 }\nonumber \\
{\mathsf  f}_4^*(\bullet\,\circ\,\circ\,\,\bullet) &=& \mathtt{010 }\nonumber \\
{\mathsf  f}_4^*(\bullet\,\circ\,\bullet\,\,\circ) &=&  \mathtt{011} \nonumber \\
{\mathsf  f}_4^*(\circ\,\circ\,\circ\,\,\bullet) &=&  \mathtt{100} \nonumber \\
{\mathsf  f}_4^*(\circ\,\circ\,\bullet\,\,\circ) &=&  \mathtt{101} \nonumber \\
{\mathsf  f}_4^*(\circ\,\bullet\,\circ\,\,\circ) &=&  \mathtt{110} \nonumber \\
{\mathsf  f}_4^*(\bullet\,\circ\,\circ\,\,\circ) &=&  \mathtt{111} \nonumber \\
{\mathsf  f}_4^*(\circ\,\circ\,\circ\,\,\circ) &=&  \mathtt{0000}. \nonumber
\end{eqnarray}
\end{example}

\medskip

We emphasize that the optimum code ${\mathsf f}_n^*$  is independent
of the design target, in that, e.g., it is the same regardless of 
whether we want to minimize average length
or the probability that the encoded length exceeds 1 KB or 1 MB.
In fact, the code ${\mathsf f}_n^*$ possesses the following
strong stochastic (competitive) 
optimality property over any other code ${\mathsf f}_n$ 
that can be losslessly decoded:
\begin{eqnarray}
\label{stop}
\mathbb{P}[ \ell ( {\mathsf f }_n (X^n )) \geq k ] \geq \mathbb{P} 
[ \ell ( {\mathsf f }_n^* (X^n )) \geq k],
\;\;\;\mbox{for all $k\geq 0$.}
\end{eqnarray}

Note that, although ${\mathsf f}_n^*$ is not a prefix code, 
the decompressor is able to recover the source file $a^n$ 
exactly from ${\mathsf f}_n^* (a^n)$ and its knowledge of 
$n$ and $P_{X^n}$. Since the whole
source file is compressed,  it is not necessary to impose 
a prefix condition in order for
the decompressor to know where the compressed file starts and ends. 
Removing the prefix-free constraint at the block level,
which is extraneous in most applications, 
results in higher compression efficiency. 

\subsection{Optimum fixed-to-variable prefix codes}
The fixed-to-variable prefix code that minimizes the average 
length is the Huffman code, achieving
the average compression rate 
$\bar{R}_{\mathsf{p}} (n)$ (which is strictly
larger than $\bar{R}(n)$),
defined as in \eqref{fl3}
but restricting the minimization to prefix codes.
Alternatively, as in \eqref{fl1}, we can investigate the 
optimum rate of the prefix code that
minimizes the probability that the length exceeds 
a given threshold. 
If the minimization in \eqref{fl1} is carried out 
with respect to codes that satisfy the prefix condition 
then the corresponding 
fundamental limit is denoted by $R_{\mathsf{p}} (n , \epsilon )$, 
and analogously $\epsilon_{\mathsf{p}}( n , k )$ for  \eqref{fl2}. Note 
that the optimum prefix code achieving the minimum in \eqref{fl2}
will, in general, depend on $k$.
The following result shows that the
corresponding fundamental limits,
with and without  the prefix condition,
are tightly coupled:

\medskip

\begin{theorem} \label{thm:prenonpre}
Suppose all elements in $\set{A}$ have positive probability. For all  $n = 1, 2, \ldots $
\begin{enumerate}
\item
For each $k=1, 2, \dots$:
\begin{eqnarray} \label{touche}
\epsilon_{\mathsf{p}}( n , k+1 ) = \left\{
\begin{array}{lr}
\epsilon^* ( n , k ) &  k < n \log_2 | \set{A} | \phantom{.}\\
0 & k \geq n \log_2 | \set{A} |.
\end{array}
\right.
\end{eqnarray}
\item If $| \set{A}| $ is not a power of 2, then for $0 \leq \epsilon < 1$: 
\begin{equation}\label{hovane}
R_{\mathsf{p}} (n , \epsilon )
= 
R^*(n,\epsilon)+\frac{1}{n}.
\end{equation}
If $| \set{A}| $ is a power of 2, then \eqref{hovane} holds for $\epsilon \geq \min_{a^n\in\set{A}^n} P_{X^n} (a^n)$, while
we have,
\begin{equation}\label{hovanes}
R_{\mathsf{p}} (n , \epsilon )
=
R^*(n,\epsilon) = \log_2 | \set{A} | + \frac{1}{n},
\end{equation}
for $0 \leq \epsilon < \min_{a^n\in\set{A}^n} P_{X^n} (a^n) $.
\end{enumerate}
\end{theorem}

\medskip

\begin{IEEEproof}
$1)\colon$ fix $k$ and $n$ satisfying $2^k < | \set{A}|^n$.
Since there is no benefit in assigning shorter lengths,
any Kraft-inequality-compliant code ${\mathsf f}_n^{\mathsf{p}}$ 
that minimizes  $\mathbb{P} [ \ell ( {\mathsf f}_n (X^n) ) >  k  ]$ 
assigns length $k$ to each of the $2^k -1$ largest masses of
$P_{X^n}$. Assigning all the other elements in $\set{A}^n$ lengths equal to
\begin{eqnarray}
\ell_{\max} = \lceil k + \log_2 ( | \set{A}|^n - 2^k + 1 ) \rceil,
\end{eqnarray}
guarantees that the Kraft sum is satisfied.
On the other hand, according to Property~1, the optimum code 
${\mathsf f}_n^*$ without prefix constraints encodes each of the 
$2^k -1$ largest masses of
$P_{X^n}$ with lengths ranging from 0 to $k - 1$.
Therefore,
\begin{eqnarray}
\mathbb{P} [ \ell ( {\mathsf f}_n^{\mathsf{p}} (X^n) ) \geq  k +1 ] 
=
\mathbb{P} [ \ell ( {\mathsf f}_n^* (X^n) ) \geq  k  ] .
\end{eqnarray}
Alternatively, if $2^k \geq | \set{A}|^n$, 
then a zero-error $n$-to-$k$ code exists, and therefore 
$\epsilon_{\mathsf{p}}( n , k+1 ) = 0$. 

$2)\colon$ According to $\mathsf{f}_n^*$ the length of the longest codeword is $ \lfloor  n \log_2 | \set{A} | \rfloor $.
Therefore, 
\begin{align} \label{safeway0}
\epsilon^*( n , \lceil n \log_2 | \set{A} | \rceil +1 ) &= 0 
\end{align}
and
\begin{align} \label{safeway00}
\epsilon^*( n , \lceil n \log_2 | \set{A} | \rceil  ) &= \left\{
\begin{array}{lr}
0 & | \set{A} | ~\mbox{is not a power of 2} \\
\min_{a^n\in\set{A}^n} P_{X^n} (a^n)  &  | \set{A} | ~\mbox{is a power of 2}
\end{array}
\right.
\end{align}
On the other hand,  1) implies
\begin{align}\label{safeway1}
\epsilon_{\mathsf{p}}( n , \lceil n \log_2 | \set{A} | \rceil +1 ) &= 0 \\
\epsilon_{\mathsf{p}}( n , \lceil n \log_2 | \set{A} | \rceil  ) &=  \epsilon^*( n , \lceil n \log_2 | \set{A} | \rceil - 1 ) \\
&\geq \min_{a^n\in\set{A}^n} P_{X^n} (a^n)\label{safeway3}
\end{align}
Furthermore, $R_{\mathsf{p}} (n , \cdot )$ can be obtained from $\epsilon_{\mathsf{p}}( n , \cdot ) $
through the counterpart to \eqref{renk}:
\begin{eqnarray} \label{renk2}
R_{\mathsf{p}}  (n, \epsilon ) = \frac{i}{n} ~~\mbox{if and only if}~ \epsilon_{\mathsf{p}} (n, i) \leq \epsilon < \epsilon_{\mathsf{p}} (n, i-1).
\end{eqnarray}
Together with \eqref{renk} and \eqref{touche}, \eqref{renk2} implies that \eqref{hovane} holds if $\epsilon \geq  \min_{a^n\in\set{A}^n} P_{X^n} (a^n) $. Otherwise, 
\eqref{safeway00}-\eqref{safeway3} result in \eqref{hovanes} when $| \set{A}|$ is a power of 2. If
$| \set{A}|$ is not a power of 2 and $0 \leq \epsilon <  \min_{a^n\in\set{A}^n} P_{X^n} (a^n) $, then
\begin{align}
R^* (n, \epsilon ) &= \frac{\lceil n \log_2 | \set{A} | \rceil}{n} \\
R_{\mathsf{p}} (n, \epsilon ) &= \frac{\lceil n \log_2 | \set{A} | \rceil +1}{n} 
\end{align}

\end{IEEEproof}

\subsection{The optimum fixed-to-fixed almost-lossless code}
As pointed out in  \cite{verduSL,verduNL}, the 
quantity
$\epsilon^* ( n , k ) $ is, in fact, intimately related 
to the problem of almost-lossless fixed-to-fixed data compression.
Assume the nontrivial compression regime in which $2^k < | \set{A}|^n$.
The optimal $n$-to-$k$ fixed-to-fixed compressor assigns a unique string of 
length $k$ to each of the $2^k - 1$ most likely  elements
of ${\cal A}^n$, and assigns all the others to the remaining binary string of 
length $k$, which signals an {\it encoder failure.}
Thus, the source strings that are decodable error-free by the optimal $n$-to-$k$ scheme
are precisely those that are encoded with 
lengths ranging from 0 to $k-1$ by the optimum 
variable-length code (Property~1).
Therefore, $\epsilon^* ( n , k ) $, defined in \eqref{fl2} as a 
fundamental limit of (strictly) lossless variable-length codes
is, in fact, equal to the minimum error probability of an $n$-to-$k$ code.
Accordingly, the results obtained in this paper apply to the 
standard paradigm of almost-lossless fixed-to-fixed compression
as well as to the setup of lossless fixed-to-variable 
compression without prefix-free constraints at the block level.

The case $2^k \geq | \set{A}|^n$ 
is rather trivial: the minimal probability
of encoding failure for an $n$-to-$k$ code is 0, 
which again coincides with $\epsilon^* ( n , k ) $,
unless
$| \set{A}|^n = 2^k$, in which case, as we saw in \eqref{safeway00},
\begin{align}
\epsilon^* ( n , k )  = \min_{a^n\in 
\set{A}^n} P_{X^n} (a^n ).
\end{align}

\subsection{Existing asymptotic results}
\label{sec:ear}
Based on the correspondence between almost-lossless fixed-to-fixed
codes and prefix-free lossless fixed-to-variable codes, 
previous results on the asymptotics of fixed-to-fixed compression 
can be brought to bear.
In particular the Shannon-MacMillan theorem  \cite{CS48,mcmillan}
implies that for a stationary ergodic finite-alphabet 
source with entropy rate $H$, and for all $0 < \epsilon <1$,
\begin{eqnarray} \label{sm}
\lim_{\ngi} R^* (n, \epsilon) = H .
\end{eqnarray}
It follows immediately from Theorem \ref{thm:prenonpre} that the prefix-free condition incurs no loss as far as the limit in \eqref{sm} is concerned:
\begin{eqnarray} \label{smprefix}
\lim_{\ngi} R_{\mathsf{p}} (n, \epsilon) = H ,
\end{eqnarray}

Suppose $X^n$ is generated by a memoryless source
with distribution
\begin{equation}
P_{X^n} = P_{X} \times P_{X}\times \cdots \times P_{X},
\end{equation}
and define the {\em information random variable},\footnote{A legacy 
of the Kraft inequality mindset,
the term ``ideal codelength" is sometimes used for
$\imath_{X} (X)$.
This is inappropriate in view 
of the fact that the optimum codelengths
are in fact {\it bounded above} by $\imath_X(X)$;
see Section~\ref{s:finite}.
Therefore, these ``ideal codelengths'' are neither ideal
nor are they actual codelengths.} 
\begin{equation} 
\label{def:info}
\imath_{X} (X) 
= \log_2\frac{1}{P_{X} (X)}.
\end{equation}
For the expected length, 
Szpankowski and Verd\'{u} \cite{wssv} show  that the behavior
of \eqref{fl3} for non-equiprobable
sources is,
\begin{eqnarray} \label{wssvg}
\bar{R}(n) = H -\frac{1}{2n } \log_2 n + O \left( \frac{1}{n} \right),
\end{eqnarray}
which is also refined to show that,
if $\imath_X(X)$ is
non-lattice,\footnote{A discrete random variable is 
{\em lattice} if all 
its masses are on a subset of some lattice
$\{ \nu + k \varsigma\;;\;k\in{\mathbb Z} \}$.} then,
\begin{eqnarray}\label{wssvnl}
\bar{R}(n) = H -\frac{1}{2n } 
\log_2 (8 \pi e \sigma^2 n ) + o \left( \frac{1}{n} \right),
\end{eqnarray}
where,
\begin{equation}
\sigma^2=\VAR(\imath_X(X)),
\label{eq:simpleVE}
\end{equation}
is the {\em varentropy} or {\em minimal coding 
variance} \cite{kontoyiannis-97} of $P_X$.
In contrast, when a prefix-free condition is imposed, 
we have the well-known behavior (see, e.g., \cite{coverthomas}),
\begin{eqnarray} \label{huffmanite}
\bar{R}_{\mathsf p} (n) = H+ O \left( \frac{1}{n} \right),
\end{eqnarray}
for any source for which 
$H=\lim_{n\to\infty}\frac{1}{n}H(X^n)$ exists.

For a non-equiprobable source such that
$\imath_X(X)$ is non-lattice,
Strassen \cite{strassen:64b} claims\footnote{See
the discussion in Section~\ref{section:memoryless}
regarding Strassen's claimed proof of this result.}
the following Gaussian approximation 
result as a refinement of \eqref{sm}:
\begin{eqnarray} \label{st}
R^* (n, \epsilon) &=&   H + \frac{\sigma}{\sqrt{n}}  Q^{-1} (\epsilon)  
-\frac{1}{2n } \log_2 \left(2 \pi \sigma^2 n e^{ ( Q^{-1} (\epsilon) )^2 }
\right) 
\nonumber
\\
&+&
\frac{\mu_3 }{6 \sigma^2 n} \left( ( Q^{-1} (\epsilon) )^2 - 1 \right) +  o\left( \frac{1}{{n}}\right).
\label{eq:Strassen}
\end{eqnarray}
Here, $Q ( x ) = \frac{1}{\sqrt{2 \pi}} \int_x^{\infty} e^{-t^2/2} \, dt $
denotes the standard Gaussian tail function, 
$\sigma^2$ is the varentropy of $P_X$
defined in (\ref{eq:simpleVE}),
and $\mu_3$ is the 
third centered absolute moment of 
the information random 
variable \eqref{def:info}.

Kontoyiannis \cite{kontoyiannis-97} 
gives a different kind of Gaussian approximation
for the codelengths $\ell({\mathsf f}_n(X^n))$
of arbitrary prefix codes ${\mathsf f}_n$ on memoryless
data $X^n$, showing that,
with probability one,
$\ell({\mathsf f}_n(X^n))$ is eventually 
lower bounded by a random variable
that has an approximately Gaussian distribution,
\begin{equation}
\ell({\mathsf f}_n(X^n))\geq Z_n
\;\;\mbox{where}\;\;
Z_n
\stackrel{\cal D}{\approx} N(nH,n\sigma^2);
\label{eq:oldCLT}
\end{equation}
and $\sigma^2$ is the varentropy as
in~(\ref{eq:simpleVE}).
Therefore, the codelengths 
$\ell({\mathsf f}_n(X^n))$ will have
at least Gaussian fluctuations
of $O(\sqrt{n})$; this
is further sharpened in \cite{kontoyiannis-97}
to a corresponding law of
the iterated logarithm,
stating that, with
probability one, the 
compressed lengths $\ell({\mathsf f}_n(X^n))$
will have fluctuations
of $O(\sqrt{n\ln \ln n})$,
infinitely often: with probability one,
\begin{equation}
\limsup_{n\to\infty}\;\frac{\ell({\mathsf f}_n(X^n))-H(X^n)}{\sqrt{2n\ln\ln n}}
\geq \sigma.
\label{eq:oldLIL}
\end{equation}
Both results (\ref{eq:oldCLT})
and (\ref{eq:oldLIL}) are 
shown to hold for Markov sources as 
well as for a wide class of mixing
sources with infinite memory.

\subsection{Outline of main new results}\label{sec:org}

Section \ref{s:finite} gives a general analysis of the distribution 
of the lengths of an optimal lossless code
for any discrete information source, which may or may not 
produce fixed-length strings of symbols. First,
in Theorems~\ref{thm:simple} and~\ref{thm:conv1}
we give simple achievability and converse bounds, 
showing that the distribution function 
of the optimal codelengths,
$\mathbb{P} \left[ \ell ( {\mathsf f }^* (X  ))  \leq t \right],$
is intimately related to the distribution of
the information random variable,
$\mathbb{P}  \left[ \imath_X (X)  \leq t \right]$.
Also we observe that the optimal codelengths 
$\ell ( {\mathsf f }^* (X  ))$ are always bounded above by
$\imath_X (X)$, but Theorem~\ref{thm:conv2} states
that they cannot be significantly smaller than
$\imath_X (X)$ with high probability.
The corresponding result for prefix codes, 
originally
derived in \cite{barronPHD,kontoyiannis-97},
is stated in Theorem~\ref{thm:converseprefix}.

Theorem~\ref{thm:viva} offers
an exact, non-asymptotic expression for 
best achievable rate $R^*(n,\epsilon)$.
So far, no other problem in information theory has 
yielded an exact non-asymptotic formula for the 
fundamental limit.
An exact expression for the average probability 
of error achieved by (almost-lossless) random binning,
is given in Theorem~\ref{thm:rb}.

General non-asymptotic and asymptotic 
results for the expected 
optimal length,
$\bar{R} (n) = 
(1/n) \mathbb{E} [\ell(\mathsf{f}_n^*(X^n))],$
are obtained in Section \ref{sec:MEL}.
Attained by the Huffman code, the minimal average length 
of prefix codes is unknown.
However, dropping the extraneous prefix constraint for 
non-symbol-by-symbol codes results in
an explicit formula for the minimal average length. 

In Section~\ref{section:asymptotic} we revisit
the refined asymptotic results (\ref{eq:oldCLT})
and (\ref{eq:oldLIL}) of \cite{kontoyiannis-97}, 
and show that they remain
valid for general (not necessarily prefix) compressors,
and for a broad class of possibly infinite-memory
sources.

Section \ref{section:memoryless} examines in detail 
the finite-blocklength behavior of the fundamental
limit $R^*(n,\epsilon)$ for the
case of memoryless sources. 
We prove tight, non-asymptotic and easily computable
bounds for $R^*(n,\epsilon)$; specifically, 
combining the results of 
Theorems~\ref{thm:iidachnonA} and~\ref{thm:RstarLB}
implies the following approximation for finite 
blocklengths $n$:

\begin{quote}
{\em Gaussian approximation I}: For every memoryless source,
the best achievable rate $R^*(n,\epsilon)$ satisfies:
\begin{equation}
nR^*(n,\epsilon)\approx nH +\sigma\sqrt{n} 
Q^{-1}(\epsilon)-\frac{1}{2}\log_2 n,
\label{eq:Gapp}
\end{equation}
where the approximation is accurate up to
$O(1)$ terms;
the same holds for 
$R_{\mathsf{p}} (n , \epsilon )$
in the case of prefix codes.
\end{quote}

The approximation~(\ref{eq:Gapp}) is established by combining the 
general results of Section \ref{s:finite}
with the classical
Berry-Ess\'{e}en bound \cite{bestBE,petrov}.
This approximation is made precise
in a non-asymptotic way, and all the constants
involved are explicitly identified.

In Section \ref{section:markov},  
achievability and converse bounds (Theorems~\ref{thm:markov-direct} and~\ref{thm:markov-converse}) are established
for $R^*(n,\epsilon)$, in the case of general 
ergodic Markov sources. Those results are analogous (though slightly weaker) to
those in Section~\ref{section:memoryless}.

We also define the varentropy rate of an arbitrary
source as the limiting normalized variance
of the information random variables $\imath_{X^n}(X^n)$, 
and we show that, for Markov chains,
it plays the same role as the varentropy defined
in (\ref{eq:simpleVE}) for memoryless sources.
Those results in particular imply the
following:

\begin{quote}
{\em Gaussian approximation II}: For any 
ergodic Markov source with entropy rate $H$ and varentropy rate $\sigma^2$,
the blocklength $n^*(R,\epsilon)$ required for 
the compression rate to exceed $(1 + \eta ) H$ 
with probability no greater than $\epsilon > 0$,
satisfies,
\begin{eqnarray}
n^* ( (1 + \eta ) H, \epsilon ) \approx 
\frac{\sigma^2}{H^2} \left( \frac{Q^{-1} (\epsilon)}{ 1 + \eta } \right)^2.
\label{eq:Gapp2}
\end{eqnarray}
[See Section~\ref{section:asymptotic} for
the general definition of the varentropy
rate $\sigma^2$, and the discussion
in Section~\ref{section:markov} for details.]
\end{quote}

Finally,
Section \ref{s:dispersion} defines the 
source dispersion $D$ as the limiting normalized 
variance of the optimal codelengths.  
In effect, the dispersion gauges the time 
one must wait for the source realization 
to become typical within a given probability,
as in (\ref{eq:Gapp2}) above, with $D$ in place
of $\sigma^2$. For a large class of sources (including 
ergodic Markov chains of any order),
the dispersion $D$ is shown to equal the 
varentropy rate $\sigma^2$ of the source.

\newpage

\section{Non-asymptotic Bounds for Arbitrary Sources}
\label{s:finite}
In this section we analyze the best achievable
compression performance on a completely general
discrete random source. In particular, (except where noted) we do not necessarily assume
that the alphabet is finite and we do not exploit the 
fact that in the original problem we are interested in compressing a
block of $n$ symbols. In this way we even encompass the case where the 
source string length is a priori unknown at the decompressor. 
Thus,  we consider a given probability mass
function $P_X$ defined
on an arbitrary finite
alphabet ${\cal X}$,
which may (but is not assumed to)
consist of variable-length strings drawn from some alphabet.
The results can then be particularized 
to the setting in Section \ref{sec:1},
letting
${\cal X} \leftarrow {\cal A}^n$ and $P_X \leftarrow P_{X^n}$.
Conversely, we can simply let $n=1$ in Section \ref{sec:1} 
to yield the setting in this section.

The best achievable rate $R^*(n,\epsilon )$ 
at blocklength $n=1$ is abbreviated as
$R^* (\epsilon )=R^*(1,\epsilon)$.
By definition, $R^*(\epsilon)$ is
the lowest $R$ such that,
\begin{eqnarray}
\mathbb{P} [  \ell ( {\mathsf f }^* (X )) > R ] \leq \epsilon, \label{samsung}
\end{eqnarray}
which is equal to the quantile function\footnote
{The quantile function ${\cal Q}\colon [0, 1] \to 
\Reals$ is the ``inverse" of the cumulative distribution 
function $F$. Specifically,
${\cal Q} ( \alpha ) = \min \{ x\colon  F ( x ) 
= \alpha \} $ if the set is nonempty; otherwise $\alpha$ lies within
a jump $\lim_{x \uparrow x_\alpha} F ( x ) < \alpha < F 
( x_\alpha )$ and we define
${\cal Q} ( \alpha ) = x_\alpha$.
}
of the integer-valued random variable $ \ell ( {\mathsf f }^* (X ))$ 
evaluated at $1 - \epsilon$.

\subsection{Achievability bound}
Recall the definition of the information
random variable $\imath_X(X)$ in~(\ref{def:info}).
Our goal is to express the distribution 
of the optimal codelengths $\ell ( {\mathsf f }^* (X ))$
in terms of the distribution of $\imath_X(X)$.
The first such result is the following simple 
and powerful upper bound (e.g. \cite{verduSL})
on the tail of the distribution of the minimum rate.

\medskip

\begin{theorem}
\label{thm:simple}
For any $a \geq 0$,
\begin{eqnarray}
\mathbb{P} \left[ \ell ( {\mathsf f }^* (X  ))  \geq a \right]  
\leq \mathbb{P}  \left[ \imath_X (X)  \geq a \right].
\end{eqnarray}
\end{theorem}
\begin{IEEEproof}
Since the labeling of the values taken by the random variable $X$ 
is immaterial, it simplifies notation in the proofs to 
assume that
the elements of $\set{X}$ are integer-valued with
decreasing probabilities: 
$P_X ( 1 ) \geq P_X (2) \geq \ldots $.  
Then, for all $i = 1, 2, \ldots $ we have the fundamental relationships:
\begin{eqnarray}
\ell ( {\mathsf f }^* ( i  )) &=& \lfloor \log_2 i \rfloor \label{lf*} \label{yom}\\
P_X (i) &\leq&  \frac{1}{i}. \label{pxi}
\end{eqnarray}
Therefore,
\begin{eqnarray}
\mathbb{P} \left[ \ell ( {\mathsf f }^* (X  ))  \geq a \right]  &=&   
\mathbb{P} \left[ \lfloor \log_2 X  \rfloor  \geq a \right]
\\
&\leq&   \mathbb{P} \left[  \log_2 X    \geq a \right]
\\
&\leq& \mathbb{P}  
\left[    \imath_X (X)   \geq a \right],
\label{wipe}
\end{eqnarray}
where \eqref{wipe} follows from \eqref{pxi}.
\end{IEEEproof}

\medskip

Before moving on, we point out that
at the core of the above proof is 
a simple but crucial observation:
not only does the distribution function 
of the optimal codelengths $\ell({\mathsf f}^*(X))$
dominate that of $\imath_X(X)$, but we in 
fact {\em always} have,
\begin{equation}
\ell({\mathsf f}^*(x))\leq
\imath_X(x),
\;\;\mbox{for all}\;x\in{\cal X}.
\label{eq:theub}
\end{equation}
This will be the used repeatedly,
throughout the rest of the paper.
Also, a simple inspection of the proof
shows that Theorem~\ref{thm:simple}
as well as~(\ref{eq:theub}) remain
valid even in the case of sources
$X$ with a countably infinite alphabet.

Theorem~\ref{thm:simple} is the 
starting point for the 
achievability result for $R^*(n,\epsilon)$
established for Markov sources in 
Theorem~\ref{thm:markov-direct}.

\subsection{Converse bounds}

In Theorem~\ref{thm:conv1}
we give a corresponding
converse result;
cf.\ \cite{verduSL}.
It will be used later to obtain
sharp converse bounds for $R^*(n,\epsilon)$
for memoryless and Markov sources,
in Theorems~\ref{thm:RstarLB}
and~\ref{thm:markov-converse},
respectively.

\medskip

\begin{theorem}\label{thm:conv1}
For any nonnegative integer $k$,
\begin{eqnarray}
\max_{{\tau} > 0} \left \{ \mathbb{P} \left[ \imath_X (X) 
	\geq k + {\tau} \right] - 2^{- {\tau} } \right\}  \leq
\mathbb{P} \left[ \ell ( {\mathsf f }^* (X  ))  \geq k \right]  .
\end{eqnarray}
\end{theorem}

\begin{IEEEproof}
As in the proof of Theorem \ref{thm:simple}, we label the values taken by 
$X$ as the positive integers in decreasing probabilities. 
Fix an arbitrary ${\tau} > 0$. Define:
\begin{eqnarray}
{\cal L} &=& \{ i \in {\cal X}\colon  {P_X (i)} \leq 2^{-k - {\tau}} \} \\
{\cal C} &=& \{ 1, 2 , \ldots 2^k  - 1 \} .
\end{eqnarray}
Then, abbreviating $P_X ( {\cal B}) = \mathbb{P} [ X \in {\cal B}] = \sum_{i \in {\cal B}} P_X (i) $, for any ${\cal B} \subset \set{X}$,
\begin{eqnarray}
\mathbb{P} \left[ \imath_X (X) \geq k + {\tau} \right] &=&  P_X ( {\cal L} )
\\
&=&  P_X ( {\cal L} \cap {\cal C} ) + P_X ( {\cal L} \cap {\cal C}^c ) 
\\
&\leq&  P_X ( {\cal L} \cap {\cal C} ) + P_X ( {\cal C}^c ) 
\\
&\leq&  ( 2^k  - 1 ) 2^{-k - {\tau}}  + P_X ( {\cal C}^c ) 
\\
&<&  2^{- {\tau}}  +  \mathbb{P} \left[ \lfloor \log_2 X \rfloor \geq k \right]
\\
&=&  2^{- {\tau}}  +  \mathbb{P} \left[ \ell ( {\mathsf f }^* (X  ))  
		\geq k \right], \label{bartok}
\end{eqnarray}
where \eqref{bartok} follows in view of \eqref{yom}.
\end{IEEEproof}

\medskip

Next we give another general converse bound, 
similar to that of Theorem~\ref{thm:conv1}, 
where this time we directly compare the 
codelengths $\ell({\mathsf f}(X))$ of an arbitrary
compressor with the values of the information 
random variable $\imath_X(X)$. Whereas
from~(\ref{eq:theub}) we know that 
$\ell({\mathsf f}(X))$ is always smaller than
$\imath_X(X)$, Theorem~\ref{thm:conv2} says
that it cannot be much smaller with high
probability. This is 
a natural analog of the corresponding 
converse established for prefix compressors
in \cite{barronPHD}, and stated
as Theorem~\ref{thm:converseprefix} below.

Applying to a finite-alphabet source, Theorem~\ref{thm:conv2} is the key
bound in the derivation of all the
pointwise asymptotic results of
Section~\ref{section:asymptotic},
Theorems~\ref{thm:SLLN}, \ref{thm:CLT}
and~\ref{thm:LIL}.
It is also the main technical
ingredient of the proof of Theorem~\ref{thm:dispersion}
in Section~\ref{s:dispersion} stating that 
the source dispersion is equal to its
varentropy.

\medskip

\begin{theorem}\label{thm:conv2}
For any compressor ${\mathsf f}$ and any $\tau>0$,
\begin{eqnarray} \label{papermaster}
\mathbb{P} \left[ \ell({\mathsf f}(X)) \leq \imath_X (X)  -\tau\right]
	\leq
	2^{-\tau} \left( \lfloor \log_2 |{\cal X}|\rfloor + 1 \right)
\end{eqnarray}
\end{theorem}

\begin{IEEEproof}
Letting ${\mathbb I}\{A\}$ denote the
indicator function of the event
$A$,
the probability in \eqref{papermaster} can be bounded by
\begin{eqnarray}
\mathbb{P} [ \ell ( \mathsf{f} (X ) ) \leq \imath_X (X) -\tau ] 
&=&
\sum_{x \in \set{X}}
P_X (x) \, {\mathbb I}
\left\{ P_X (x) \leq 2^{-\tau-  \ell ( \mathsf{f} (x ) )}\right\} \\
&\leq&
 2^{-\tau} \sum_{x \in \set{X}} 2^{-  \ell ( \mathsf{f} (x ) )},
\label{eq:nonKraft}
\\
&\leq& 2^{-\tau} \sum_{j=0}^{\lfloor \log_2 |{\cal X}|\rfloor} 2^j 2^{-j}\label{warmoun}
\end{eqnarray}
where  the sum in~(\ref{eq:nonKraft}) 
is maximized if $\mathsf{f}$ assigns a string of length $j+1$
only if it also assigns all strings of length $j$.
Therefore, \eqref{warmoun} holds because that code contains
all the strings of lengths $0, 1, \ldots, \lfloor \log_2 |{\cal X}|\rfloor - 1$
plus $|{\cal X}| - 2^{\lfloor \log_2 |{\cal X}|\rfloor} + 1 \leq 2^{\lfloor \log_2 |{\cal X}|\rfloor}$ strings of length $\lfloor \log_2 |{\cal X}|\rfloor$.
\end{IEEEproof}
We saw in Theorem \ref{thm:prenonpre}
that the optimum prefix code under the criterion 
of minimum excess length probability incurs a penalty of 
at most one bit. The following elementary converse 
is derived in \cite{barronPHD,kontoyiannis-97};
its short proof is included for completeness.
Indeed, the statements and proofs of Theorems~\ref{thm:conv2} and \ref{thm:converseprefix}
are close parallels.

\medskip

\begin{theorem}\label{thm:converseprefix}
For any prefix code $\mathsf{f}$, and any $\tau \geq 0$:
\begin{eqnarray}
\mathbb{P} [ \ell ( \mathsf{f} (X ) ) 
< \imath_X (X) -\tau ] \leq 2^{-\tau}.
\end{eqnarray}
\end{theorem}

\medskip

\begin{IEEEproof} We have, as
in the proof of Theorem~\ref{thm:conv2}
leading to (\ref{eq:nonKraft}),
\begin{eqnarray}
\mathbb{P} [ \ell ( \mathsf{f} (X ) ) < \imath_X (X) -\tau ] 
&<&
2^{-\tau} \sum_{x \in \set{X}} 2^{-  \ell ( \mathsf{f} (x ) )}  \\
&\leq& 2^{-\tau}, \label{kin}
\end{eqnarray}
where \eqref{kin} is Kraft's inequality. 
\end{IEEEproof}

%

\subsection{Exact fundamental limit}
The following result expresses the 
non-asymptotic data compression fundamental limit 
$R^*(\epsilon)=R^*(1,\epsilon)$
as a function of the source information spectrum.

\medskip

\begin{theorem} \label{thm:viva}
For all $a \geq 0$, the exact minimum rate compatible with 
given excess-length probability  satisfies,
\begin{eqnarray} \label{apple}
R^* ( \epsilon ) = \left \lceil \log_2 
\left( 1 + M(2^a) \right) \right \rceil - 1,
\end{eqnarray}
with,
\begin{eqnarray}
\epsilon = \mathbb{P} [ \imath_X ( X ) \geq a ], \label{centim}
\end{eqnarray}
where $M(\beta)$ denotes the number of masses with 
probability strictly larger than $\frac{1}{\beta}$, 
and which can be expressed as:
\begin{eqnarray} \label{comb2prob}
M(\beta) = \beta \,
\mathbb{P} \left[ \imath_X (X) < \log_2 \beta \right]
- \int_1^\beta
\mathbb{P} \left[ \imath_X (X) \leq \log_2 t \right] \, dt.
\end{eqnarray}
\end{theorem}

\begin{IEEEproof}
As above, the values taken by 
$X$ are labeled as the positive integers 
in order of decreasing probability. 
By the definition of $M(\cdot)$, for any positive integer $i$, and $a>0$,
\begin{eqnarray}
P_X (i) \leq 2^{-a} &\Longleftrightarrow&
\log_2 ( 1 + M(2^a) )
 \leq \log_2 i,
\end{eqnarray}
and it is easy to check that:
\begin{eqnarray}
\lceil \alpha \rceil - 1 <   \lfloor \log_2 i \rfloor  &\Longleftrightarrow& \alpha \leq  \log_2 i.
\end{eqnarray}

Therefore, letting $\alpha =  \log_2 ( 1 + M(2^a) ) $ and letting the integer-valued $X$ take the role of $i$, we obtain 
that \eqref{samsung} is satisfied with equality if $R$ is given by the 
right side of \eqref{apple}.
Any smaller value of $R$ would prevent \eqref{samsung} from being satisfied.
\par
The proof of \eqref{comb2prob}  follows a sequence of elementary steps:
\begin{eqnarray}
M (\beta) &=& \sum_{x \in \set{X} } 
{\mathbb I}\left\{ P_X (x) > \frac{1}{\beta} \right\} \\
&=& 
\mathbb{E} 
\left[
\frac{{\mathbb I} \{ P_X (X) > \frac{1}{\beta} \}}{P_X(X)}
\right]
\\
&=&
\int_0^{\infty}
\mathbb{P}
\left[
\frac{{\mathbb I} \{ P_X (X) > \frac{1}{\beta} \}}{P_X(X)} > t \right] \, dt
\\
&=&
\int_0^{\beta}
\mathbb{P}
\left[
\frac{1}{\beta} < P_X(X) <  \frac{1}{t} \right] \, dt
\\&=&
\int_0^{\beta}
\mathbb{P}
\left[
\frac{1}{\beta} <    P_X(X) \right]
-
\mathbb{P}
\left[
 P_X(X)  \geq \frac{1}{t} \right] 
   \, dt
\\&=&
\beta \,
\mathbb{P} \left[ \imath_X (X) < \log_2 \beta \right]
-
\int_1^\beta
\mathbb{P} \left[ \imath_X (X) \leq \log_2 t \right] \, dt .
\end{eqnarray}
\end{IEEEproof}

\medskip

While Theorem \ref{thm:viva}  
gives $R^*(\epsilon)=R^*(1,\epsilon)$ exactly for those 
$\epsilon$ which correspond to values taken by the 
complementary cumulative distribution function
of the information random variable $\imath_X (X)$,
a continuous sweep of $a>0$ gives a very dense grid of values,
unless $X$ (whose alphabet size typically
grows exponentially with $n$ in the 
fixed-to-variable setup)
takes values in a
very small alphabet.
From the value of $a$ we can obtain the probability in the
right side of 
\eqref{centim}. The optimum code achieves that excess probability 
$\epsilon = \mathbb{P} \left[ \ell ( {\mathsf f }^* (X  ))  \geq \ell_a \right]  $
for lengths equal to,
\begin{eqnarray}
\ell_a = \lceil a + \log_2 ( 2^{-a} + 2^{-a} M ( 2^a ) ) \rceil,
\end{eqnarray}
where the second term is negative and represents the exact gain with respect to
the information spectrum of the source.

For later use we observe that, 
if we let $M_X^+(\beta)$ be the number of masses 
with probability  larger or equal than $\frac{1}{\beta}$, 
then,\footnote{Where typographically convenient we use $\exp ( a ) = 2^a$.}
\begin{eqnarray} \label{liolla}
M_X^+ (\beta) 
&=& 
\sum_{x \in \set{X} } 
{\mathbb I}\left\{ P_X (x) \geq \frac{1}{\beta} \right\} \\
&=& \mathbb{E} 
\left[
	\exp \left( \imath_X (X) \right) 
	{\mathbb I}\left\{ \imath_X (X) \leq \log_2\beta \right\}
\right].
\end{eqnarray}

\medskip

\begin{figure}[ht]
\centerline{\includegraphics[width=12cm]{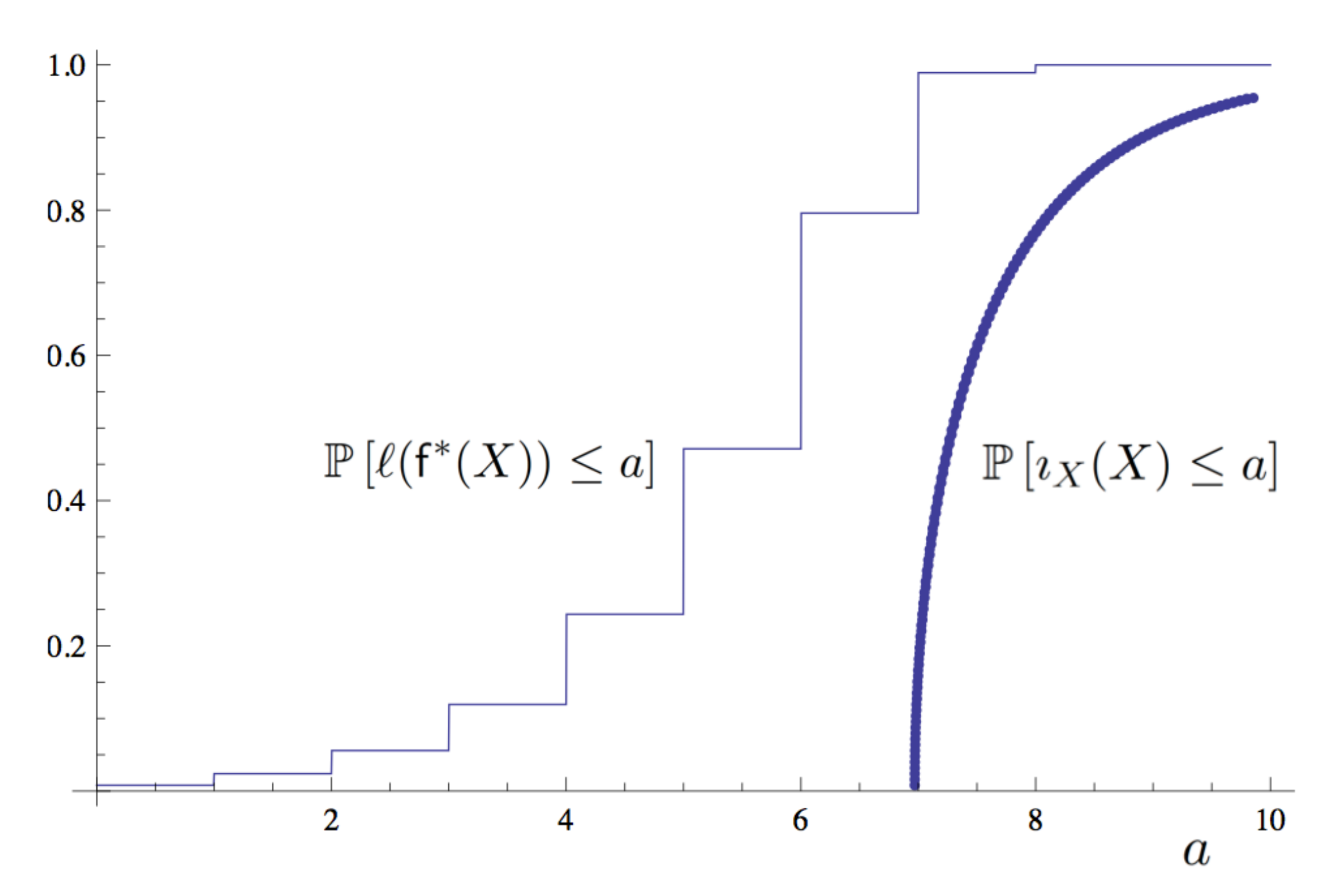}}
\caption{\label{fig:binomial} Cumulative distribution functions 
of $\ell ( {\mathsf f }^* (X  ))$ and $\imath_X (X)$ 
when $X$ is the number of tails obtained in 10,000 fair coin flips.}
\end{figure}

Figure~\ref{fig:binomial}
shows the cumulative
distribution functions of  $\ell ( {\mathsf f }^* (X  ))$ and $\imath_X (X)$ 
when $X$ is a binomially distributed random variable: 
the number of tails obtained in 10,000 fair coin flips. 
Therefore, $\imath_X (X)$ ranges from $6.97 
\approx 10,000 - \log_2 \binom{10000}{5000}$ to $10,000$ and,
\begin{eqnarray}
H (X) = 7.69 \\
\mathbb{E} [\ell ( {\mathsf f }^* (X  ))] = 6.29,
\end{eqnarray}
where all figures are in bits.

\subsection{Exact behavior of random binning}
The following result 
gives an exact expression for the 
performance of random binning for arbitrary sources,
as a function of the 
cumulative distribution function 
of the random variable $\imath_X (X)$
via \eqref{comb2prob}.
In binning, the compressor is no 
longer constrained to be an injective mapping.
When the label received by the decompressor can be explained by  more than
one source realization, it chooses the most likely one,
breaking ties arbitrarily.
(Cf.\ \cite{PPV} for the exact 
performance of random coding in channel coding.)

\medskip

\begin{theorem}\label{thm:rb}
Averaging uniformly over all binning compressors 
${\mathsf f}\colon  {\cal X} \to \{1,2,\ldots N\}$, 
results in an expected 
error probability equal to,
\begin{eqnarray}
1 - \mathbb{E}
\left[
\sum_{\ell = 0}^{J(X)-1} 
\frac{
\binom{J(X)-1}{\ell}
}
{N^\ell (1+\ell)}
\left( 1 - \frac{1}{N}\right)^{M( \frac{1}{P_X (X)})+ J(X) - \ell - 1} 
\right],
\label{exex}
\end{eqnarray}
where $M(\cdot)$ is
given in \eqref{comb2prob}, and the number of masses whose 
probability is equal to $P_X (x) $ is denoted by:
\begin{eqnarray}
J(x) = \frac{\mathbb{P} [ P_X (X) = P_X (x) ]}{P_X (x)}.
\end{eqnarray}
\end{theorem}

\medskip

\begin{IEEEproof}
For the purposes of the proof, it is convenient to assume 
that ties are broken uniformly at random among the most
likely source outcomes in the bin.
To verify \eqref{exex}, note that,
given that the source realization is 
$x_0$:
\begin{enumerate}
\item
The number of masses with probability strictly higher than that of $x_0$ is
$M ( \frac{1}{P_X (x_0)})$;
\item
Correct decompression of $x_0$ requires that any $x$ with $P_X (x) > P_X (x_0)$ not be assigned
to the same bin as $x_0$. This occurs with probability:
\begin{eqnarray}
\left( 1 - \frac{1}{N}\right)^{M( \frac{1}{P_X (x_0)})};
\end{eqnarray}
\item
If there are $\ell$ masses with the same probability as $x_0$ in the same bin,
correct decompression occurs with probability $\frac{1}{1+\ell}$.
\item
The probability that there are $\ell$ masses with the same probability as $x_0$ in the same
bin is equal to:
\begin{eqnarray}
\binom{J(x_0)-1}{\ell}
\left( 1 - \frac{1}{N}\right)^{ J(x_0) - \ell - 1} \frac{1}{N^\ell}.
\end{eqnarray}
\end{enumerate}

Then, \eqref{exex} follows since all the bin assignments are independent.
\end{IEEEproof}

\medskip

Theorem \ref{thm:rb} leads to an achievability bound for
both almost-lossless fixed-to-fixed compression
and lossless fixed-to-variable compression.
However, in view of the simplicity and tightness 
of Theorem \ref{thm:simple}, the main usefulness
of Theorem \ref{thm:rb}
is to gauge the suboptimality of random binning 
in the finite (in fact, rather short because of computational complexity)
blocklength regime.

\newpage

\section{Minimal Expected Length}\label{sec:MEL}
Recall the definition of 
the best achievable rate $\bar{R}(n)$ in Section~\ref{sec:1},
expressed in terms of 
$\mathsf{f}_n^*$ as in~(\ref{eq:Rbarn}).
An immediate consequence of Theorem \ref{thm:simple} is the bound,
\begin{eqnarray} \label{nydk}
n\bar{R}(n)=
\mathbb{E} \left[ \ell ( \mathsf{f}^* ( X ) )\right] \leq  H(X),
\end{eqnarray}
which goes back at least to the work of Wyner  \cite{wyner}.
Indeed, by lifting the prefix condition it is possible to beat the 
entropy on average as we saw in the asymptotic results \eqref{wssvg} 
and \eqref{wssvnl}.
Lower bounds on the minimal average length
as a function of $H(X)$ can be found in \cite{wssv} and references therein.
An explicit expression can be obtained easily 
by labeling the outcomes as the positive
integers with decreasing probabilities as in the proof of 
Theorem \ref{thm:simple}:
\begin{eqnarray}
\mathbb{E}  [ \ell ( {\mathsf  f }^* (X  ))  ] &=& \mathbb{E} [ \lfloor \log_2 X \rfloor ] \\
&=&
\sum_{k=1}^\infty \mathbb{P} [ \lfloor \log_2 X \rfloor  \geq k] \\
&=&
\sum_{k=1}^\infty \mathbb{P} [   X   \geq 2^k]. \label{tspaty}
\end{eqnarray}
\begin{example}
The average number of bits required to encode at which flip of a fair coin the first tail appears is
equal to,
\begin{eqnarray}
\sum_{k=1}^\infty \mathbb{P} [   X   \geq 2^k] &=& \sum_{k=1}^\infty \sum_{j=2^k}^\infty 
2^{-j}
\\
&=&
2 \sum_{k=1}^\infty 2^{2^{-k}}
\\
&\approx& 
0.632843,
\end{eqnarray}
since, in this case, $X$ is a geometric random variable with $ \mathbb{P} [   X = j ] = 2^{-j}$.
In contrast, imposing a prefix constraint disables any compression: the optimal prefix code consists of all,
possibly empty, strings of $\texttt{0}$s terminated by $\texttt{1}$, achieving an average length of 2.
\end{example}

\medskip

\begin{example}\label{example:equiprobable}
If $X_M$ is equiprobable on a set of $M$ elements, then:
\begin{enumerate}
\item
\begin{eqnarray}
\mathbb{E} \left[ \ell ( {\mathsf  f }^* ( X_M  )) \right]
=
 \lfloor \log_2 M \rfloor 
+   \frac{1}{M} \left( 
 2 + \lfloor \log_2 M \rfloor  - 2^{\lfloor \log_2 M \rfloor + 1}
 \right),
\label{humongous}
\end{eqnarray}
which simplifies to,
\begin{eqnarray} \label{boy}
\mathbb{E} \left[ \ell ( {\mathsf  f }^* ( X_M  )) \right] =
\frac{(M+1) \log_2 (M+1)}{M} - 2,
\end{eqnarray}
when $M+1$ is a power of 2.
\item
\begin{eqnarray}
\limsup_{M \rightarrow \infty} H(X_M) - \mathbb{E} \left[ \ell ( {\mathsf  f }^* ( X_M  )) \right] &=& 2 \\
\liminf_{M \rightarrow \infty} H(X_M) -  \mathbb{E} \left[ \ell ( {\mathsf  f }^* ( X_M  )) \right] &=& 1 + \log_2 e - \log_2 \log_2 e,
\end{eqnarray}
where the entropy is expressed in bits.
\end{enumerate}
\end{example}

\medskip

\medskip

\begin{theorem}\label{thm:expgen}
For any source $\mathbf{X} = \{P_{X^n}\}_{n=1}^\infty$ with finite 
entropy rate,
\begin{eqnarray}
H ( \mathbf{X} ) = \limsup_{n \to \infty} \frac{1}{n} H (X^n) < \infty,
\label{eq:Hrate}
\end{eqnarray}
the normalized minimal average length satisfies:
\begin{eqnarray} \label{dkny}
\limsup_{n \to \infty} \bar{R} (n) = H ( \mathbf{X} ).
\end{eqnarray}
\end{theorem}
\begin{IEEEproof}
The achievability (upper) bound in \eqref{dkny} holds in view of \eqref{nydk}.
In the reverse direction, we invoke the bound \cite{AO94}:
\begin{eqnarray} \label{alonorlobo}
 H(X^n)  - \mathbb{E}  [ \ell ( {\mathsf  f }_n^* (X^n  ))  ] 
\leq
 \log_2 (H(X^n) + 1) +\log_2 e.
\end{eqnarray}
Upon dividing both sides of \eqref{alonorlobo} by $n$ and taking $\limsup$ the desired
result follows, since for any $\delta >0$, for all
sufficiently large $n$,
$H(X^n) \leq n H ( \mathbf{X} ) + n \delta$.
\end{IEEEproof}

\medskip

In view of \eqref{huffmanite},
we see that
the penalty incurred on the \textit{average rate} 
by the prefix condition vanishes asymptotically
in the very wide generality allowed by Theorem \ref{thm:expgen}. 
In fact, the same proof we used for Theorem \ref{thm:expgen}
shows the following result:

\medskip

\begin{theorem} \label{thm:meanratio}
 For any (not necessarily serial) source $\mathbf{X} = \{P_{X^{(n)}}\}_{n=1}^\infty$,
\begin{eqnarray}
\lim_{n \to \infty}
\frac{\bar{R}(n) }{\frac{1}{n}H ({X^{(n)}} )} 
= 
\lim_{n \to \infty}
\frac{ \mathbb{E}  [ \ell ( {\mathsf  f }_n^* (X^{(n)}  ))  ]}{H ({X^{(n)}} )} 
= 1,
\end{eqnarray}
as long as $H(X^{(n)})$ diverges, 
where $X^{(n)} \in \set{A}_n$, an alphabet 
which is not necessarily a Cartesian product.
\end{theorem}

\newpage

\section{Pointwise Asymptotics}\label{section:asymptotic}
\subsection{Normalized pointwise redundancy}
Before turning to the precise evaluation of the
best achievable rate $R^*(n,\epsilon)$,
in this section we examine the asymptotic behavior of the normalized difference between the
codelength and the information (sometimes known as the pointwise redundancy). 

\medskip

\begin{theorem}\label{thm:sickSV}
For any discrete source and any
divergent deterministic sequence $\kappa_n$ such that,
\begin{equation}
\lim_{n \to \infty} \frac{\log n}{\kappa_n} = 0,
\end{equation}
we have:
\begin{description}
\item[(a)]
For any sequence $\{{\mathsf f}_n\}$ of codes:
\begin{equation}
\liminf_{n\to\infty}\frac{1}{\kappa_n} \left( \ell({\mathsf f}_n(X^n) )- \imath_{X^n} (X^n) \right ) \geq 0,
\;\;\;\mbox{w.p.1.} 
\label{sickSVa}
\end{equation}
\item[(b)]
The sequence of optimal codes $\{{\mathsf f}^*_n\}$ achieves:
\begin{equation}
\liminf_{n\to\infty}\frac{1}{\kappa_n} \left( \ell({\mathsf f}^*_n(X^n) )- \imath_{X^n} (X^n) \right ) = 0,
\;\;\;\mbox{w.p.1.} 
\label{sickSVb}
\end{equation}
\end{description}
\end{theorem}
\begin{IEEEproof}
(a) We invoke the general
converse in Theorem~\ref{thm:conv2}, with $X^n$ and ${\cal A}^n$ 
in place of $X$ and ${\cal X}$, respectively.
Fixing arbitrary $\epsilon > 0$ and letting $\tau=\tau_n= \epsilon \kappa_n$, we obtain that,
\begin{equation}\label{artesian}
\mathbb{P} \left[ \ell({\mathsf f}_n(X^n)) \leq \imath_{X^n} (X^n)- \epsilon \kappa_n\right]
	\leq
	 2^{\log_2 n -\epsilon \kappa_n}  \left( \log_2 |{\cal A}|+1 \right)
\end{equation}
which is summable in $n$. Therefore, the Borel-Cantelli lemma implies that the $\limsup$ of the event on the left side of \eqref{artesian}
has zero probability, or equivalently, with probability one,  
\[  { \ell({\mathsf f}_n(X^n)) - \imath_{X^n} (X^n)} \geq - \epsilon { \kappa_n}\]
is violated only a finite number of times. Since $\epsilon$ can be chosen arbitrarily small, \eqref{sickSVa} follows.
\indent
Part~(b) follows from (a) and  (\ref{eq:theub}).
\end{IEEEproof}
\subsection{Stationary Ergodic Sources}
 Theorem~\ref{thm:expgen} states  that for any discrete process $\mathbf{X}$
the expected rate of the optimal codes
${\mathsf f}^*_n$ satisfy,
\begin{eqnarray}
\limsup_{n \to \infty} \frac{1}{n} \mathbb{E}[\ell({\mathsf f}_n^*(X^n))] = H (\mathbf{X}).
\end{eqnarray}
The next result shows that if the source
is stationary and ergodic, then the same asymptotic
relation holds not just in expectation,
but also with probability 1. Moreover,
no compressor can beat the entropy rate asymptotically
with positive probability.
The corresponding results for prefix codes 
were established in \cite{barronPHD,kieffer:91,kontoyiannis-97}. 

\medskip

\begin{theorem}
\label{thm:SLLN}
Suppose that $\{X_n\}$ is a stationary ergodic source
with entropy rate $H$.
\begin{itemize}
\item[(i)]
For any sequence $\{{\mathsf f}_n\}$ of codes,
\begin{equation}
\liminf_{n\to\infty}\frac{1}{n}\ell({\mathsf f}_n(X^n))\geq H,
\;\;\;\mbox{w.p.1.} 
\label{eq:LLNd}
\end{equation}
\item[(ii)]
The sequence of optimal codes $\{{\mathsf f}^*_n\}$ achieves,
\begin{equation}
\lim_{n\to\infty}\frac{1}{n}\ell({\mathsf f}^*_n(X^n)) = H,
\;\;\;\mbox{w.p.1.}
\end{equation}
\end{itemize}
\end{theorem}

\begin{IEEEproof}
The Shannon-Macmillan-Breiman theorem states that,
\begin{eqnarray}
\frac{1}{n} \imath_{X^n} (X^n) \to H, \;\;\;\mbox{w.p.1.}
\end{eqnarray}
Therefore, the result is an immediate consequence of Theorem \ref{thm:sickSV} with $\kappa_n = n$.
\end{IEEEproof}

\subsection{Stationary Ergodic Markov Sources}
We assume that the
source is a stationary ergodic 
(first-order) Markov chain, with transition
kernel,
\begin{equation}
P_{X'|X} ( x'\,|\,x )
\;\;\;\;(x,x')\in \set{A}^2,
\end{equation}
on the finite alphabet ${\cal A}$.
Further restricting the source to be Markov enables us to analyze more precisely the behavior of the information random variables
and, in particular, we will show that the zero-mean random variables,
\begin{eqnarray}
Z_n =  \frac{\imath_{X^n} ( X^n ) - H(X^n) }{\sqrt{n}},
\end{eqnarray}
are asymptotically normal with variance given by the varentropy rate, which generalizes the notion in
\eqref{eq:simpleVE}.

\medskip

\begin{definition}
The varentropy rate of a random process $\mathbf{X} = \{ P_{X^n} \}_{n=1}^\infty$ is 
\begin{equation}
\sigma^2\;=\;\limsup_{n\to\infty}\,\frac{1}{n}\,
\VAR(\imath_{X^n}(X^n)).
\label{eq:ve}
\end{equation}
\end{definition}

\medskip

\noindent
Some remarks are in order:
\begin{itemize}
\item If $\mathbf{X}$ is a stationary memoryless process each of whose letters is distributed according to $P_{\mathsf{X}}$,
then the varentropy rate of $\mathbf{X}$ is equal to the varentropy of ${\mathsf{X}}$. The varentropy of ${\mathsf{X}}$
is zero if and only if it is equiprobable on its support.
\item
In contrast to the first moment, we do not know whether stationarity is sufficient for $\limsup = \liminf$ in \eqref{eq:ve}.
\item
While the entropy-rate of a Markov chain admits a two-letter expression,
the varentropy does not. In particular, if $\sigma^2  ( a ) $ denotes the varentropy of the distribution $P_{X'|X} ( \cdot\,|\,a )$, then the varentropy of the chain is, in general, not given by $ \mathbb{E} [ \sigma^2  ( X_0 ) ]$.
\item
The varentropy rate of 
Markov sources is typically nonzero.
For example, for a first order
Markov chain it was observed in
\cite{yushkevich,kontoyiannis-jtp} that
$\sigma^2=0$ if and only if 
the source satisfies the following
\textit{deterministic equipartition property}:
Every  string $x^{n+1}$ that 
starts and ends with the same symbol,
has probability 
(given that $X_1=x_1$)
$q^n$, 
for some constant $0 \leq q \leq 1$. 
\end{itemize}

\medskip

\begin{theorem}
\label{thm:CLT}
Let $\{X_n\}$ be a stationary ergodic finite-state Markov chain.
\begin{itemize}
\item[(i)] The varentropy rate $\sigma^2$ is also
equal to the corresponding $\liminf$
of the normalized variances in~(\ref{eq:ve}),
and it is finite.
\item[(ii)]
The normalized information random variables 
are asymptotically normal, in the sense that,
as $n\to\infty$,
\begin{equation}\label{jijiji}
 \frac{\imath_{X^n} ( X^n ) -  H(X^n) }{\sqrt{n}} 
\longrightarrow N (0, \sigma^2 ),
\end{equation}
in distribution.
\item[(iii)]
The normalized information random variables 
satisfy a corresponding law of the iterated logarithm:
\begin{align}\label{sssuuuppp}
\limsup_{n \to \infty} \frac{\imath_{X^n} ( X^n ) -  H(X^n) }{\sqrt{2 n \ln \ln n}} &= \sigma,\;\;\;\mbox{w.p.1}
\\
\label{iiinnnfff}
\liminf_{n \to \infty} \frac{\imath_{X^n} ( X^n ) -  H(X^n) }{\sqrt{2 n \ln \ln n}} &= - \sigma,\;\;\;\mbox{w.p.1}
\end{align}
\end{itemize}
\end{theorem}
\begin{IEEEproof}

(i) and (ii): Consider the bivariate Markov chain
$\{\tilde{X}_n=(X_n,X_{n+1})\}$
on the alphabet 
${\cal B}=\{(x,y)\in {\cal A}^2 \colon P_{X'|X}(y|x)>0\}$
and  the function
$f\colon {\cal B}\to{\mathbb R}$ defined by
\begin{equation}
f(x,y)= \imath_{X'|X}(y|x).
\end{equation}
Since $\{X_n\}$ is stationary and ergodic,
so is $\{\tilde{X}_n\}$, hence, by 
the central limit theorem for functions of Markov chains \cite{chung:book} 
\begin{eqnarray} \label{sumthu}
\frac{1}{\sqrt{n}} \left( \imath_{X^n|X_1} (X^n | X_1) - H(X^n|X_1) \right) = 
\frac{1}{\sqrt{n}} \sum_{i=1}^{n-1}\,(f(\tilde{X}_i)-{\mathbb E}[ f(\tilde{X}_i)])
\end{eqnarray}
converges in distribution to the zero-mean Gaussian law with finite variance
\begin{equation}
\Sigma^2\;=\;\lim_{n\to\infty}\,\frac{1}{n}
\VAR(\imath_{X^n|X_1}(X^n|X_1)).
\end{equation}
Furthermore, since
\begin{eqnarray} \label{peaches}
\imath_{X^n}(X^n)-H(X^n) = \imath_{X^n|X_1}(X^n|X_1) - H(X^n|X_1) + \left( \imath_{X_1}(X_1) - H(X_1)\right)
\end{eqnarray}
where the second term is bounded,   \eqref{jijiji} must hold 
and we must have $\Sigma^2 = \sigma^2$.

(iii) Normalizing \eqref{sumthu} by $\sqrt{2 n \ln \ln n}$ in lieu of $\sqrt{n}$, we can invoke the law of the iterated logarithm for functions of Markov chains \cite{chung:book} to show that the $\limsup/\liminf$ of the sum 
behave as claimed. Since upon normalization, the second term in the right side of \eqref{peaches}, vanishes almost surely,
$\imath_{X^n}(X^n)-H(X^n)$ must satisfy the same behavior.
%
%
%
\end{IEEEproof}

\medskip
Together with Theorem \ref{thm:sickSV} particularized to $\kappa_n = \sqrt{n}$,
we conclude that the normalized deviation of the optimal codelengths from the entropy rate 
satisfies
\begin{equation}\label{jojojo}
 \frac{\ell (\mathsf{f}_n^* ( X^n ) ) -  H }{\sqrt{n}} \longrightarrow N (0, \sigma^2 )
\end{equation}
which is the same behavior as that exhibited by the Shannon prefix code \cite{kontoyiannis-97},
so as far as the pointwise $\sqrt{n}$ asymptotics the prefix constraint does not entail loss of efficiency.
Similarly, the following result readily follows from 
Theorem \ref{thm:CLT} and Theorem \ref{thm:sickSV} with $\kappa_n = \sqrt{2 n \ln \ln n}$.
\medskip

\begin{theorem}
\label{thm:LIL}
Suppose $\{X_n\}$ is a stationary ergodic
Markov chain with entropy rate $H$ and varentropy
rate $\sigma^2$. Then:
\begin{itemize}
\item[(i)]
For any sequence of codes $\{{\mathsf f}_n\}$:
\begin{eqnarray}
&&
\limsup_{n\to\infty}\;\frac{\ell({\mathsf f}_n(X^n))-H(X^n)}{\sqrt{2n\ln\ln n}}
\geq \sigma,\;\;\;\mbox{w.p.1;}
\label{eq:lil1}\\
&&
\liminf_{n\to\infty}\;\frac{\ell({\mathsf f}_n(X^n))-H(X^n)}{\sqrt{2n\ln\ln n}}
\geq -\sigma,\;\;\;\mbox{w.p.1.}
\label{eq:lil2}
\end{eqnarray}
\item[(ii)]
The sequence of optimal codes $\{{\mathsf f}_n^*\}$ 
achieves the bounds in~(\ref{eq:lil1})
and (\ref{eq:lil2}) with equality.
\end{itemize}
\end{theorem}

\medskip

The Markov sufficient condition in Theorem \ref{thm:CLT}
enabled the application of the central limit theorem and the law of the iterated logarithm
to the sum in \eqref{sumthu}. According to Theorem~9.1 of \cite{philipp-stout:book} a more general sufficient condition is that
$\{X_n\}$ be a stationary process with
$\alpha(d) = O(d^{-336})$ and $\gamma(d) = O(d^{-48}),$
with the
mixing coefficients:
\begin{align}
\gamma(d)&=\max_{a\in {\cal A}}{\mathbb E}\,\Big|
\imath_{X_0 | X_{-\infty}^{-1}} ( a | X_{-1}, X_{-2},\ldots )
-
\imath_{X_0 | X_{-d}^{-1}} ( a | X_{-1}, X_{-2},\ldots  X_{-d})
\Big|\\
\alpha(d)&=\sup\left\{|{\mathbb P}(B\cap A)-
{\mathbb P}(B){\mathbb P}(A)|\;;\;
A\in{\cal F}_{-\infty}^0,\;B\in{\cal F}_{d}^{\infty}\right\}.
\end{align}
Here ${\cal F}_{-\infty}^0$ 
and ${\cal F}_{d}^{\infty}$ denote the 
$\sigma$-algebras generated by the collections of
random variables $(X_0,X_{-1},\ldots)$ and $(X_d,X_{d+1},\ldots)$,
respectively.
The $\alpha(d)$ are the {\em strong mixing} coefficients 
\cite{bradley:86} of $\{X_n\}$, and
the $\gamma(d)$ were introduced by
Ibragimov in \cite{ibragimov:62}.
Although 
these mixing conditions 
may be hard to verify in practice,
they are fairly weak in that they 
require only polynomial decay of 
$\alpha(d)$ and $\gamma(d).$
In particular, any ergodic Markov chain of 
any order satisfies these conditions.

\newpage

\section{Gaussian Approximation for Memoryless Sources}
\label{section:memoryless}
We turn our attention to the non-asymptotic
behavior of the best rate $R^*(n,\epsilon)$
that can be achieved when compressing 
a stationary memoryless finite-alphabet source
$\{X_n \in \set{A}\}$ with marginal distribution $P_{X}$, 
whose entropy and varentropy are denoted 
by $H$ and $\sigma^2$, respectively.

Specifically, we will derive explicit upper and lower
bounds on $R^* (n , \epsilon )$ in terms
of the first three moments of the information 
random variable $\imath_{X} (X)$.
Although particularizing Theorem~\ref{thm:viva} 
it is possible, in principle, to compute $R^*(n,\epsilon)$
exactly,
it is more desirable to derive approximations
that are both easier to compute and offer more intuition 
into the behavior of 
the fundamental limit $R^*(n,\epsilon)$.

Theorems~\ref{thm:iidachnonA}
and~\ref{thm:RstarLB}
imply that, for all $\epsilon\in(0,1/2)$,
the best achievable rate 
$R^*(n,\epsilon)$ satisfies,
\begin{equation}
\frac{c}{n}
\leq
R^*(n,\epsilon)-
\left[ 
	H+\frac{\sigma}{\sqrt{n}} Q^{-1}(\epsilon)-\frac{\log_2 n}{2 n} 
\right]
\leq \frac{c'}{n}.
\label{eq:iidG}
\end{equation}
The upper bound is valid for all $n$,
the lower bound is valid for $n\geq n_0$
as in~(\ref{nmino}), and explicit
values are derived for
the constants $c,c'$.
In view of Theorem~\ref{thm:prenonpre},
essentially the same results as in~(\ref{eq:iidG})
hold for prefix codes, 
\begin{equation}
\frac{c}{n}
\leq
R_{\mathsf p}(n,\epsilon)-
\left[ 
	H+\frac{\sigma}{\sqrt{n}} Q^{-1}(\epsilon)-\frac{\log_2 n}{2 n} 
\right]
\leq \frac{c'+1}{n}.
\label{eq:iidG2}
\end{equation}
The bounds in equations~(\ref{eq:iidG})
and~(\ref{eq:iidG2}) justify 
the Gaussian approximation~(\ref{eq:Gapp})
stated in Section~\ref{sec:1}.


Before establishing the precise non-asymptotic
relations leading to~(\ref{eq:iidG})
and~(\ref{eq:iidG2}), we illustrate 
their utility via an example.
To facilitate this, note that 
Theorem~\ref{thm:simple} 
immediately yields the following simple bound:

\medskip

\begin{theorem}
For all $n\geq 1$, $\epsilon>0$,
\begin{eqnarray} \label{rune}
R^* (n , \epsilon ) \leq R^{\mathsf{u}} (n , \epsilon),
\end{eqnarray}
where $R^{\mathsf{u}} (n , \epsilon)$ is the quantile 
function of the information spectrum, i.e., the lowest $R$ such that:
\begin{eqnarray}
\mathbb{P} \left[ \frac{1}{n} \sum_{i=1}^n 
\imath_{X} (X_i)  \geq R \right] \leq \epsilon.
\end{eqnarray}
\end{theorem}

\medskip

\begin{figure}[ht]
\centerline{\includegraphics[width=15cm]{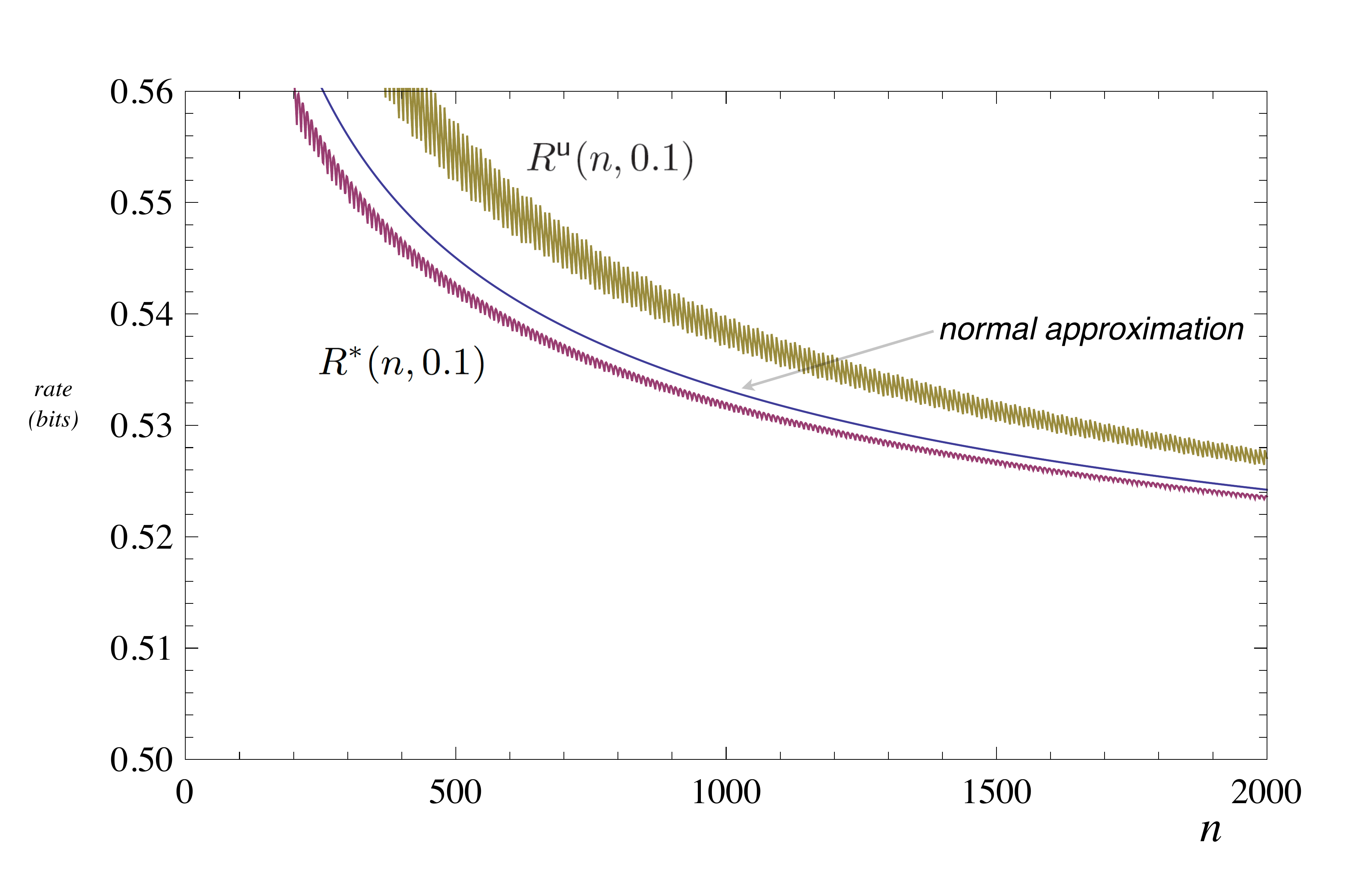}}
\caption{\label{fig:2000}
The optimum rate $R^*(n, 0.1)$, the
Gaussian approximation 
$\tilde{R}^*(n, 0.1)$
in \eqref{normal3},
and the upper bound
$R^{\mathsf{u}} (n , 0.1)$,
for a Bernoulli-$0.11$ source 
and blocklengths $200\leq n\leq 2000$.}
\end{figure}

In Figure  \ref{fig:2000},  we exhibit the 
behavior of the 
fundamental compression limit $R^*(n,\epsilon)$
in the case of coin flips with bias $0.11$
(for which $H \approx 0.5$ bits). 
In particular, we compare $R^*(n, \epsilon)$ and 
$R^{\mathsf{u}} (n, \epsilon)$ for $\epsilon = 0.1$. 
The non-monotonic nature of both $R^*(n, \epsilon)$ and 
$R^{\mathsf{u}} (n, \epsilon)$ with $n$ is not surprising:
although the larger the value of $n$ the less we are at the 
mercy of the source randomness, we also need to compress 
more information.
Figure \ref{fig:2000} also illustrates
that $R^*(n, \epsilon)$ is tracked rather closely by
the Gaussian approximation,
\begin{eqnarray}\label{normal3}
\tilde{R}^*(n, \epsilon) = H + Q^{-1} ( \epsilon ) 
\frac{\sigma}{\sqrt{n}} - \frac{1}{2n} \log_2 n,
\end{eqnarray}
suggested by~(\ref{eq:iidG}).

\begin{figure}[ht]
\centerline{\includegraphics[width=15cm]{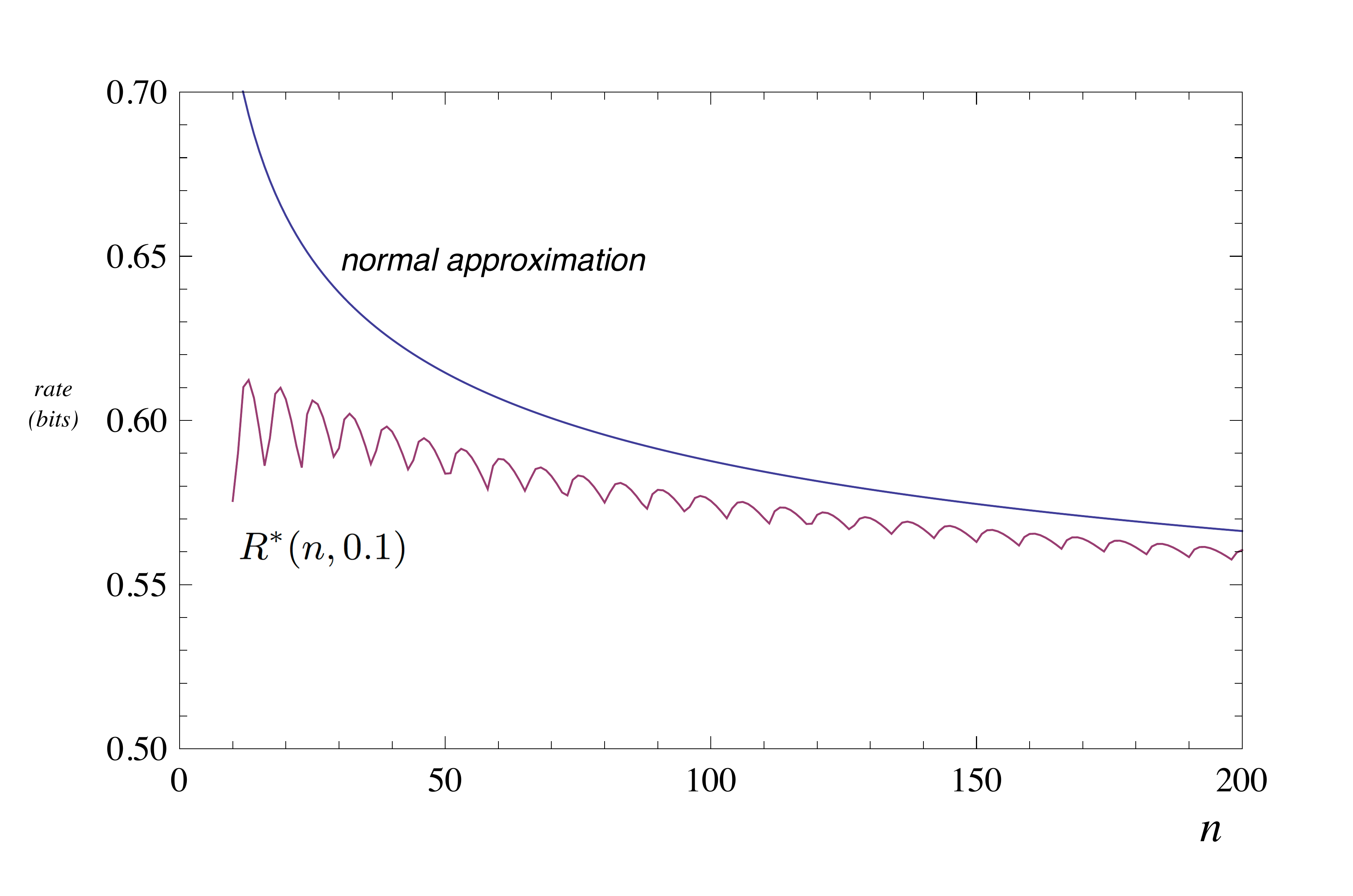}}
\caption{\label{fig:200} 
The optimum rate $R^*(n, 0.1)$ and the Gaussian approximation 
$\tilde{R}^*(n, 0.1)$  in \eqref{normal3}, for a Bernoulli-$0.11$ source
and blocklengths $10\leq n\leq 200$.}
\end{figure}

Figure \ref{fig:200} focuses the comparison between ${R}^*(n, 0.1)$ 
and $\tilde{R}^*(n, 0.1)$ on the short blocklength range up to $200$ 
not shown in  Figure  \ref{fig:2000}. For $n > 60$, the discrepancy 
between the two never exceeds 4\%.

The remainder of the section is devoted to justifying 
the use of 
\eqref{normal3} as 
an accurate approximation to $R^* (n , \epsilon )$.
To that end, 
in Theorems~\ref{thm:RstarLB} and~\ref{thm:iidachnonA}
we establish the bounds given in~(\ref{eq:iidG}).
Their derivation 
requires that we overcome two technical hurdles: 
\begin{enumerate}
\item
The distribution function of the optimal 
encoding length is not the same as the
distribution of  $\frac{1}{n} \sum_{i=1}^n \imath_{X} (X_i) $;
\item
The distribution of $\frac{1}{n} \sum_{i=1}^n \imath_{X} (X_i) $ 
is only approximately Gaussian.
\end{enumerate}

\medskip

To cope with the second hurdle we will appeal to the 
classical Berry-Ess\'{e}en bound \cite{bestBE}, \cite{petrov}:

\medskip

\begin{theorem}\label{thm:BE} 
Let $\{Z_i\}$ be independent and identically distributed
random variables with zero mean and unit variance, and
let $\bar{Z}$ be standard normal.
Then, for all $n\geq 1$ and any $a$:
\begin{align} \label{lejos}
\left| \mathbb{P} \left[  \frac{1}{\sqrt{n}} \sum_{i=1}^n Z_i \leq a \right] 
- \mathbb{P} \left[ \bar{Z} \leq a \right] \right| \leq \frac{\mathbb{E} [ |Z_1 -\mathbb{E}[Z_1]|^3]}{2 \sqrt{n}} .
\end{align}
\end{theorem}

\medskip

Invoking the Berry-Ess\'{e}en bound, Strassen \cite{strassen:64b}
claimed the following approximation for $n > \frac{19600}{\delta^{16} }$,
\begin{eqnarray} \label{stnona}
\left| R^* (n, \epsilon) -   \tilde{R}^*(n, \epsilon) \right| 
\leq \frac{140}{\delta^8},
\end{eqnarray}
where
\begin{align}
\delta &\leq \min\left\{\sigma, \epsilon, 1- \epsilon, \mu_3^{-1/3} \right\} \\
\mu_3  &=  \mathbb{E} [ |\imath_{X} (X) - H |^3 ].
\end{align}
Unfortunately, we were not able to verify how \cite{strassen:64b} justifies 
the application of \eqref{lejos}  to bound integrals with respect to the 
corresponding cumulative distribution functions (cf.\ equations
(2.17), (3.18) and the displayed equation between (3.15) and (3.16)
in \cite{strassen:64b}).

The following achievability result holds for all blocklengths.

\medskip

\begin{theorem}\label{thm:iidachnonA}
For all $0 < \epsilon \leq \frac{1}{2}$ and all $n \geq 1$,
\begin{eqnarray}
R^*(n,\epsilon) 
&\leq& 
H+\frac{\sigma}{\sqrt{n}}  Q^{-1} (\epsilon)
-\frac{\log_2 n}{2n } \nonumber\\
&&+\frac{1}{n} \log_2  \left( \frac{\log_2 e}{ \sqrt{2 \pi \sigma^2} } 
+  \frac{\mu_3}{\sigma^3} \right)\nonumber\\
&&+
\frac{1}{n} 
\frac {\mu_3}{\sigma^2 \phi(\Phi^{-1}(\Phi(Q^{-1}(\epsilon))
+\frac{\mu_3}{\sigma^3\sqrt{n}}))},  
\end{eqnarray}
as long as the varentropy $\sigma^2$ is strictly positive,
where $\Phi = 1 - Q$ and $\phi=\Phi'$ are the
standard Gaussian distribution function and density,
respectively.
\end{theorem}

\medskip

\begin{IEEEproof}
The proof follows Strassen's construction, but 
the essential approximation steps are different.
The positive constant $\beta_n$ is uniquely defined by:
\begin{eqnarray} \label{kovan}
\mathbb{P} \left[ \imath_{X^n} (X^n)\leq\log_2 \beta_n\right]&\geq& 1-\epsilon,
\label{eq:mu1}\\
\mathbb{P} \left[ \imath_{X^n} (X^n) <\log_2 \beta_n\right]	 &<& 	1-\epsilon.
\label{eq:mu2}
\end{eqnarray}
Since the information spectrum (i.e., the distribution 
function of the information random variable
$\imath_{X^n} (X^n)$) is piecewise constant,  
$\log_2 \beta_n$ is the location of the jump where the 
information spectrum reaches (or exceeds for the first time) 
the value $1-\epsilon$.
Furthermore, defining the normalized constant,
\begin{eqnarray}
\lambda_n=\frac{\log_2\beta_n-nH}{\sqrt{n}\sigma},
\label{eq:lambdan}
\end{eqnarray}
the probability in the left side of \eqref{kovan} is,
\begin{eqnarray} \label{pw1}
\mathbb{P} \left[ \frac{\imath_{X^n} (X^n) - n H}{\sqrt{n} \sigma } \leq \lambda_n \right]
\leq \Phi ( \lambda_n) + \frac{\mu_3}{2 \sigma^3 \sqrt{n}},
\end{eqnarray}
where we have applied Theorem \ref{thm:BE}.
Analogously, we obtain,
\begin{eqnarray}\label{pw2}
\mathbb{P} \left[ \frac{\imath_{X^n} (X^n) - n H}{\sqrt{n} \sigma } < \lambda_n \right]
\geq \Phi ( \lambda_n) - \frac{\mu_3}{2 \sigma^3 \sqrt{n}}.
\end{eqnarray}
Since $1- \epsilon$ is sandwiched between the right sides 
of \eqref{pw1} and \eqref{pw2}, 
as $n\to\infty$ we must have that, 
$\lambda_n \rightarrow \lambda$,
where,
\begin{eqnarray}
\lambda = \Phi^{-1} ( 1- \epsilon ) = Q^{-1} (\epsilon ).
\end{eqnarray}
By a simple first-order Taylor bound,
\begin{eqnarray}
\lambda_n
&\leq&
	\Phi^{-1}\Big(\Phi(\lambda)+\frac{\mu_3}{2 \sigma^3 \sqrt{n}}\Big)
	\\
&=&
	\lambda+\frac{\mu_3}{2 \sigma^3\sqrt{n}}(\Phi^{-1})'(\xi_n )
	\\
&=&
	\lambda+\frac{\mu_3}{2 \sigma^3 \sqrt{n}}\frac{1}{
	\phi(\Phi^{-1}(\xi_n ))},
	\label{eq:lambdaTaylor}
\end{eqnarray}
for some $\xi_n \in[\Phi(\lambda),\Phi(\lambda)+\frac{\mu_3}{2 \sigma^3\sqrt{n}}]$.
Since $\epsilon\leq 1/2$, we have $\lambda\geq0$ and
$\Phi(\lambda)\geq1/2$, so that $\xi_n \geq 1/2$.
And since $\Phi^{-1}(t)$ is strictly increasing for all $t$, 
while $\phi$ is strictly decreasing for $t\geq 0$, 
from (\ref{eq:lambdaTaylor}) we obtain,
\begin{eqnarray}
\lambda_n
\leq\lambda+\frac{\mu_3}{2 \sigma^3 \sqrt{n}}
\frac{1}{\phi(\Phi^{-1}(\Phi(\lambda)+\frac{\mu_3}{2 \sigma^3\sqrt{n}}))}.
\label{gio}
\end{eqnarray}

The event  $E_n$ in the left
side of \eqref{kovan} contains all the ``high probability strings,''
and itself it has probability $\geq 1-\epsilon$.  Its cardinality
is $ M_X^+ ( \beta_n )$, defined in \eqref{liolla} (with $X \leftarrow X^n$).
Therefore, denoting,
\begin{eqnarray}
p(t) &=& 2^{-t} {\mathbb I} \{ t \geq 0 \} \\
Y_i &=& \frac{1}{\sigma} \left( \imath_{X} ( X_i ) - H \right),
\label{abnormals}
\end{eqnarray}
we obtain,
\begin{eqnarray}
R^*(n,\epsilon)
&\leq& \frac{1}{n} \log_2 M_X^+ ( \beta_n ) \\
&=&
 \frac{1}{n} \log_2 \mathbb{E} 
\left[
	\exp \left( \imath_{X^n} ({X^n}) \right) 
	{1}\left\{ \imath_{X^n} ({X^n}) \leq \log_2 \beta_n \right\}
\right]
 \\
&=&
H + \lambda_n \frac{\sigma}{\sqrt{n}}
+ \frac{1}{n} \log_2 \alpha_n, \label{vanni}
\end{eqnarray}
with,
\begin{eqnarray}
\alpha_n &= & \mathbb{E}
\left[ 
\, p ( \log_2 \beta_n - \imath_{X^n} ({X^n})  )
\right] \\
&=&
 \mathbb{E}
\left[ 
p \left( \sqrt{n} \sigma \left( \lambda_n - \frac{1}{\sqrt{n}} \sum_{i=1}^n Y_i  \right) \right)
\right],
\label{cranny}
\end{eqnarray}
and where the $\{ Y_i \}$ are independent, identically distributed, 
with zero mean and unit variance.
Let $\bar{\alpha}_n$ be defined as \eqref{cranny} except that $Y_i$ are replaced by
$\bar{Y}_i$ which are standard normal.
Then, straightforward algebra yields,
\begin{eqnarray}
\bar{\alpha}_n &=& \mathbb{E} \left[
2^{-\sqrt{n} \sigma ( \lambda_n - \bar{Y}_1 ) } 
{\mathbb I} \{ \bar{Y}_1 \leq \lambda_n \}
\right]
\\
&=&
\int_0^{\infty} 2^{-x} \frac{e^{-\frac{( x + \lambda_n \sigma \sqrt{n})^2}{2 \sigma^2 n}}}{ \sqrt{2 \pi \sigma^2 n} } \, dx
\\
&\leq&
\frac{\log_2 e}{ \sqrt{2 \pi \sigma^2 n} }.
\end{eqnarray}
To deal with the fact that 
the random variables in \eqref{cranny} are not normal,
we apply the Lebesgue-Stieltjes integration 
by parts formula to \eqref{cranny}.
Denoting the distribution of the normalized sum in 
\eqref{cranny} by $F_n(t)$,
$\alpha_n$ becomes,
\begin{eqnarray}
 \alpha_n &=& \int_{-\infty}^{\lambda_n} 
2^{- \left( \sqrt{n} \sigma \left( \lambda_n - t  \right) \right)}
dF_n (t) \\
&=&
 F_n (\lambda_n)  - \int_{-\infty}^{\lambda_n}
F_n (t) \sqrt{n} \sigma 2^{- \left( \sqrt{n} \sigma \left( \lambda_n - t  \right) \right)} dt \log_e 2
 \\
&=&
\bar{\alpha}_n +   F_n (\lambda_n)  - \Phi (\lambda_n) 
\nonumber \\
&&- \int_{-\infty}^{\lambda_n}
(F_n (t) - \Phi (t) )\sqrt{n} \sigma 2^{- \left( \sqrt{n} \sigma \left( \lambda_n - t  \right) \right)} dt \log_e 2
\\
&\leq&
\bar{\alpha}_n + \frac{\mu_3 }{2 \sigma^3\sqrt{n}}
 + \frac{ \mu_3}{2 \sigma^2}
\int_{-\infty}^{\lambda_n}
2^{- \left( \sqrt{n} \sigma \left( \lambda_n - t  \right) \right)} dt \log_e 2
\label{kew}\\
&=&
\bar{\alpha}_n +\frac{ \mu_3}{ \sigma^3 \sqrt{n} } 
\\
&\leq&
\frac{1}{\sqrt{n}} \left( \frac{\log_2 e}{ \sqrt{2 \pi \sigma^2} } 
+ \frac{ \mu_3}{\sigma^3} \right),
\label{crannyL}
\end{eqnarray}
where \eqref{kew} follows from Theorem \ref{thm:BE}.
The desired result now follows from \eqref{vanni} 
after assembling the bounds on $\lambda_n$ and $\alpha_n$ in 
\eqref{gio} and \eqref{crannyL}, respectively.
\end{IEEEproof}

\medskip

Next we give a complementary converse result.

\medskip

\begin{theorem}
\label{thm:RstarLB}
For all $0 < \epsilon < \frac12$ and all $n$ such that,
\begin{eqnarray} \label{nmino}
n > 
n_0=\frac14 \left( 1 + \frac{\mu_3}{2 \sigma^3} \right)^2
\frac{1}{\left( \phi ( Q^{-1}  ( \epsilon) )Q^{-1} ( \epsilon)\right)^2},
\end{eqnarray}
the following lower bound holds,
\begin{eqnarray}
R^*(n,\epsilon)
\geq 
	H+\frac{\sigma}{\sqrt{n}} Q^{-1}(\epsilon)-\frac{\log_2 n}{2 n} 
	-\frac{\frac{\mu_3}{2} +\sigma^3 }{n \sigma^2 \phi(Q^{-1}(\epsilon))},
\label{eq:thm4}
\end{eqnarray}
as long as the varentropy $\sigma^2$ is strictly positive.
\end{theorem}

\medskip

\begin{IEEEproof}
Let,
\begin{eqnarray}
\eta = \frac{\frac{\mu_3}{2 \sigma^2} +\sigma }{ \phi(Q^{-1}(\epsilon))},
\end{eqnarray}
and consider,
\begin{eqnarray} 
&&\mathbb{P} \left[ \sum_{i=1}^n \imath_{X} (X_i) 
	\geq H n  +  \sigma \sqrt{n} Q^{-1}(\epsilon) - \eta \right]
\nonumber\\
&=&
\mathbb{P} \left[ \sum_{i=1}^n \frac{\imath_{X} (X_i) - H}{\sigma \sqrt{n}}
	\geq   Q^{-1}(\epsilon) - \frac{\eta}{ \sigma \sqrt{n}} \right]\label{correr} \\
&\geq&
Q \left( Q^{-1}(\epsilon) - \frac{\eta}{ \sigma \sqrt{n}} \right) - \frac{\mu_3}{2 \sigma^3 \sqrt{n}} \label{alfredo}
\\
&\geq&
\epsilon + \frac{\eta}{ \sigma \sqrt{n}} \phi (  Q^{-1}(\epsilon) ) - \frac{\mu_3}{2 \sigma^3 \sqrt{n}} \label{germont}
\\
&=&
\epsilon + \frac{ 1 }{ \sqrt{n}}, \label{webnet}
\end{eqnarray}
where \eqref{alfredo}
follows from Theorem \ref{thm:BE}, and 
\eqref{germont} follows from,
\begin{eqnarray}
Q( a  - \Delta ) \geq Q (a) + \Delta \phi ( Q(a) ),
\end{eqnarray}
which holds at least as long as,
\begin{eqnarray}
a > \frac{\Delta}{2} > 0. \label{pasato}
\end{eqnarray}
Letting $a = Q^{-1}(\epsilon)$ and $\Delta = \frac{\eta}{ \sigma \sqrt{n}}$,
\eqref{pasato} is equivalent to \eqref{nmino}.

We proceed to invoke Theorem~\ref{thm:conv1} with $X \leftarrow X^n$, 
$k$ equal to $n$ times the right side of \eqref{eq:thm4},
and $\tau = \frac{1}{2} \log_2 n$.
In view of the definition of $R^* (n, \epsilon)$ 
and \eqref{correr}-\eqref{webnet},
the desired result follows.
\end{IEEEproof}

\newpage

\section{Gaussian Approximation for Markov Sources}
\label{section:markov}

Let $\{X_n\}$ be an irreducible, 
aperiodic, $k$th order Markov chain
on the finite alphabet ${\cal A}$,
with transition probabilities,
\begin{eqnarray}
P_{X'|X^k} ( x_{k+1}\,|\,x^k ),
\;\;\;\;x^{k+1}\in \set{A}^{k+1},
\end{eqnarray}
and entropy rate $H$.
Note that we do not assume that
the source is stationary.
In Theorem~\ref{thm:CLT} of 
Section~\ref{section:asymptotic}
we established that the varentropy rate
defined in general in equation~(\ref{eq:ve}),
for stationary ergodic chains exists as the
limit,
\begin{equation}
\sigma^2\;=\;\lim_{n\to\infty}\,\frac{1}{n}\,
\VAR(\imath_{X^n}(X^n)).
\label{eq:vel}
\end{equation}
An examination of the proof shows that,
by an application of the general
central limit theorem for (uniformly 
ergodic) Markov chains \cite{chung:book,meyn-tweedie:book2},
the assumption
of stationarity is not necessary,
and~(\ref{eq:vel}) holds for all 
irreducible aperiodic chains.

\medskip

\begin{theorem}
\label{thm:markov-direct} 
Suppose $\{X_n\}$ is an
irreducible and aperiodic
$k$th order Markov source, and let
$\epsilon\in(0,1/2)$. 
Then, there is a positive constant $C$ 
such that, for all $n$ large enough,
\begin{eqnarray}
nR^*(n,\epsilon) \leq nH+\sigma\sqrt{n}Q^{-1}(\epsilon)+C,
\label{eq:directM}
\end{eqnarray}
where the varentropy
rate $\sigma^2$ is given
by (\ref{eq:vel}) and it
is assumed to be strictly positive.
\end{theorem}

\medskip

\begin{theorem}
\label{thm:markov-converse}
Under the same assumptions as in Theorem~\ref{thm:markov-direct},
for all $n$ large enough,
\begin{eqnarray}
n R^*(n,\epsilon)
\geq nH+\sigma\sqrt{n}Q^{-1}(\epsilon)-\frac{1}{2}\log_2 n
	-C,
\label{eq:thm6}
\end{eqnarray}
where $C>0$ is a finite constant, 
possibly different from than in Theorem~\ref{thm:markov-direct}.
\end{theorem}

\medskip

\noindent
{\em Remarks. }
\begin{enumerate}
\item
By definition, the lower bound in Theorem \ref{thm:markov-converse} also applies to $R_{\mathsf p}(n,\epsilon)$,
while in view Theorem \ref{thm:prenonpre}, the upper bound in Theorem \ref{thm:markov-direct} also
applies to $R_{\mathsf p}(n,\epsilon)$ provided $C$ is replaced by $C+1$.
\item
Note that, unlike the direct and converse
coding theorems
for memoryless
sources (Theorems~\ref{thm:iidachnonA} 
and~\ref{thm:RstarLB}, respectively)
the results of Theorems~\ref{thm:markov-direct} 
and~\ref{thm:markov-converse}
are asymptotic in
that we do not give explicit bounds
for the constant terms.
This is because the main probabilistic
tool we use in the proofs (the 
Berry-Ess\'{e}en bound in Theorem~\ref{thm:BE}) does not 
have an equally precise counterpart
for Markov chains. Specifically,
in the proof of Theorem~\ref{thm:MarkovBE} below 
we appeal to a Berry-Ess\'{e}en bound 
established by Nagaev in \cite{nagaev:61},
which does not give an explicit value for
the multiplicative constant $A$ in~(\ref{eq:MarkovBE}).
More explicit
bounds do exist, but they require additional
conditions on the Markov chain;
see, e.g., Mann's thesis \cite{mann:phd}, and the references
therein. 
\item
If we restrict our attention to the 
(much more narrow) class of {\em reversible}
chains, then it is indeed possible to apply the
Berry-Ess\'{e}en bound of Mann \cite{mann:phd} to obtain
explicit values for the constants
in Theorems~\ref{thm:markov-direct} and~\ref{thm:markov-converse};
but the resulting values are pretty loose, 
drastically limiting the engineering usefulness
of the resulting bounds.
For example, in Mann's version of the
Berry-Ess\'{e}en bound, the corresponding
right side of the inequality as in
Theorem~\ref{thm:BE} is multiplied by a factor
of 13000. Therefore, we have opted for the less
explicit but much more general statements given
above.
\item
Similar comments to those in the last two remarks
apply to the observation that
Theorem~\ref{thm:markov-direct} is a weaker
bound than that established in Theorem~\ref{thm:iidachnonA} 
for memoryless sources, by a $(1/2)\log_2 n$ term.
Instead of restricting our result to the much more narrow 
class of reversible chains, or extending the 
involved proof of Theorem~\ref{thm:iidachnonA}
to the case of a Markovian source, we chose to illustrate
how this slightly weaker bound can be established in 
full generality, with a much shorter and simpler proof.
\item 
The proof of Theorem~\ref{thm:markov-direct}
shows that the constant in its statement can be chosen as
\begin{equation}\label{thechosen}
C=\frac{2A\sigma}{\phi(Q^{-1}(\epsilon))}
\end{equation}
for all 
\begin{equation}\label{tribe}
n\geq
\frac{8A^2}
{\pi e(\phi(Q^{-1}(\epsilon)))^4},
\end{equation}
where $A$ is the constant appearing in Theorem~\ref{thm:MarkovBE},
below.
Similarly, from the proof of Theorem~\ref{thm:markov-converse}
we see that
the constant in its statement can be chosen as
\begin{eqnarray}
C = \frac{\sigma(A+1)}{\phi(Q^{-1}(\epsilon))}+1,
\end{eqnarray}
for all,
\begin{equation}
n\geq\left(
\frac{A+1}{Q^{-1}(\epsilon)\phi(Q^{-1}(\epsilon))}
\right)^2.
\end{equation}
Note that, in both cases, the values of the constants
can easily be improved, but they still depend on the implicit constant $A$ 
of Theorem~\ref{thm:MarkovBE}.
\end{enumerate}

\medskip

As mentioned above, we will need 
a Berry-Ess\'{e}en-type bound on the scaled 
information random variables,
\[\frac{\imath_{X^n}(X^n) - nH}{\sqrt{n} \sigma}.\]
Beyond the
Shannon-McMillan-Breiman theorem,
several more refined asymptotic
results have been established for
this sequence; see, in particular,
\cite{ibragimov:62,philipp-stout:book,strassen:64b,yushkevich}
and the discussions in \cite{kontoyiannis-jtp}
and in Section~\ref{section:asymptotic}.
Unlike these asymptotic results, we will
use the following non-asymptotic bound.

\medskip

\begin{theorem} 
\label{thm:MarkovBE} 
For an ergodic, $k$th order Markov source $\{X_n\}$
with entropy rate $H$ and positive varentropy rate $\sigma^2$, there exists
a finite constant $A>0$ such that, for all $n\geq 1$,
\begin{eqnarray}
\sup_{z\in\Reals}
\left|
\mathbb{P}\Big[\imath_{X^n}(X^n)-nH> z \,{\sigma\sqrt{n}}\Big]
-Q(z)\right|\leq\frac{A}{\sqrt{n}}.
\label{eq:MarkovBE}
\end{eqnarray}
\end{theorem}

\begin{IEEEproof}
For integers $i\leq j$, 
we adopt the notation $x_i^j$ and $X_i^j$ for blocks
of strings 
$(x_i,x_{i+1},\ldots,x_j)$ and random
variables 
$(X_i,X_{i+1},\ldots,X_j)$, respectively.
For all
$x^{n+k}\in\set{A}^{n+k}$ such that $P_{X^k}(x^k) >0 $
and $P_{X'| X_{j-k}^{j-1}} (x_j\,|\,x_{j-k}^{j-1}) >0$, 
for $j=k+1, k+2,\ldots n+k$, we have,
\begin{eqnarray}
\imath_{X^n}(x^n)
&=&
	\log_2\frac{1}{P_{X^k}(x^k)\prod_{j=k+1}^n
	P_{X'| X_{j-k}^{j-1}}(x_j\,|\,x_{j-k}^{j-1})}
	\\
&=&
	\sum_{j=k+1}^{k+n}\log_2 \frac{1}{P_{X'| X_{j-k}^{j-1}}(x_j\,|\,x_{j-k}^{j-1})}
	\nonumber\\
&-&
	\log_2
	\frac{P_{X^k}(x^k)}{\prod_{j=n+1}^{n+k}P_{X'| X_{j-k}^{j-1}}(x_j\,|\,x_{j-k}^{j-1})}
	\label{eq:temp}
	\\
	&=&
		\sum_{j=1}^n f(x^{j+k})
	\;+\;
	\Delta_n,
	\label{eq:Delta1}
\end{eqnarray}
where the function $f\colon A'\to\Reals$ is defined by,
\begin{eqnarray}
f(x^{k+1})= \imath_{X'|X^k} ( x_{k+1} | x^k ) = \log_2 \frac{1}{P_{X'|X^k} ( x_{k+1}\,|\,x^k )},
\end{eqnarray}
and,
\begin{eqnarray}
\Delta_n = -
	\log_2
	\frac{P_{X^k}(x^k)}{\prod_{j=n+1}^{n+k}
	P_{X'| X_{j-k}^{j-1}}(x_j\,|\,x_{j-k}^{j-1})}.
\end{eqnarray}
Denote
\begin{eqnarray}
|\Delta_n|\leq\delta = \max
\left|
	\log_2
	\Big[
	\frac{P_{X^k}(x^k)}{\prod_{j=n+1}^{n+k}
	P_{X'|X^k}(x_j\,|\,x_{j-k}^{j-1})}
	\Big]
\right|<\infty,
\label{eq:Delta2}
\end{eqnarray}
where the maximum is over 
the positive probability strings for which we have 
established \eqref{eq:Delta1}.

Let $\{Y_n\}$ denote the first-order Markov
source defined by taking overlapping
$(k+1)$-blocks in the original chain, 
\begin{eqnarray}
Y_n=(X_n,X_{n+1},\ldots,X_{n+k}).
\end{eqnarray}
Since $\{X_n\}$ is irreducible
and aperiodic, so is $\{Y_n\}$ on the
state space, 
\begin{eqnarray}
\set{A}'=\{x^{k+1}\in \set{A}^{k+1}\colon ~P_{X'|X^k} (x_{k+1}|x^k)>0\}.
\end{eqnarray}
Now, since the chain $\{Y_n\}$
is irreducible and aperiodic on a finite
state space, condition~(0.2) of
\cite{nagaev:61} is satisfied, and
since the function $f$ is bounded, 
Theorem~1 of \cite{nagaev:61} implies
that there exists a
finite constant $A_1$ such that,
for all $n$,
\begin{eqnarray}
\sup_{z\in\Reals}
\left|
\mathbb{P}\Big[\frac{\sum_{j=1}^n f(Y_j)-nH}{\sigma\sqrt{n}} > z\Big]
-Q(z)\right|\leq\frac{A_1}{\sqrt{n}},
\end{eqnarray}
where the entropy rate is $H = \mathbb{E} [ f (\tilde{Y}_1)]$
and, 
\begin{eqnarray}
\Sigma^2=
\lim_{n\to\infty}\frac{1}{n} \mathbb{E}\Big[
\Big(\sum_{j=1}^n(f(\tilde{Y}_j)-H)\Big)^2
\Big],
\label{eq:nagaev}
\end{eqnarray}
where $\{\tilde{Y}_n\}$
is a stationary version of $\{Y_n\}$,
that is, it has the same transition
probabilities but its initial distribution
is its unique invariant distribution,
\begin{equation}
\mathbb{P} [\tilde{Y}_1=x^{k+1}]=\pi(x^k)P_{X'|X^k}(x_{k+1}\,|\,x^k),
\end{equation}
where $\pi$ is the unique invariant distribution
of the original chain $\{X_n\}$.
Since the function $f$ is bounded and the
distribution of the chain $\{Y_n\}$
converges to stationarity exponentially 
fast, it is easy to see that \eqref{eq:nagaev} coincides with the source 
varentropy rate.

Let $F_n(z)$, $G_n(z)$ denote the complementary
cumulative distribution functions, 
\begin{eqnarray}
F_n (z) &=& \mathbb{P} \left[ \imath_{X^n}(X^n) - nH
	> z {\sqrt{n} \sigma}\right], \\
G_n (z) &=&  \mathbb{P} \left[ \sum_{j=1}^nf(Y_j)-nH 
	> z {\sqrt{n}\sigma}\right].
\end{eqnarray}
Since  $F_n(z)$ and $G_n(z)$ are non-increasing, (\ref{eq:Delta1}) and
(\ref{eq:Delta2}) imply that
\begin{eqnarray}
F_n(z)
&\geq& 
G_n(z+\delta/\sqrt{n})\\
&\geq&
Q(z+\delta/\sqrt{n})-\frac{A_1}{\sqrt{n}}, \label{himo}
\\
&\geq&
Q(z)-\frac{A}{\sqrt{n}},
\label{eq:halfU}
\end{eqnarray}
uniformly in $z$,
where \eqref{himo} follows from
(\ref{eq:nagaev}), and \eqref{eq:halfU}
holds with $A=A_1+\delta/\sqrt{2\pi}$ since
$Q'(z)=-\phi(z)$ is bounded by $- 1/\sqrt{2\pi}$.
A similar argument shows that,
\begin{eqnarray}
F_n(z)
&\leq& 
G_n(z-\delta/\sqrt{n})\\
&\leq&
Q(z-\delta/\sqrt{n})+\frac{A_1}{\sqrt{n}}\\
&\leq&
Q(z)+\frac{A}{\sqrt{n}}.
\label{eq:halfL}
\end{eqnarray}
Since both (\ref{eq:halfU}) and (\ref{eq:halfL})
hold uniformly in $z\in\Reals$, together 
they form the statement of the theorem.
\end{IEEEproof}

\medskip

\begin{IEEEproof}[Proof of Theorem~\ref{thm:markov-direct}]
Starting from
Theorem~\ref{thm:simple}
with $X^n$ 
in place of $X$ and with,
\begin{eqnarray}
K_n=nH+\sigma\sqrt{n}Q^{-1}(\epsilon)+C,
\end{eqnarray}
where $C$ will be chosen below,
Theorem~\ref{thm:simple}
states that,
\begin{eqnarray}
	\mathbb{P}[\ell({\mathsf f}_n^*(X^n)) \geq K_n ]
&\leq&
	\mathbb{P}[\imath_{X^n}(X^n)
	\geq K_n
	]\\
&=&
	\mathbb{P}\left[ \frac{1}{\sigma\sqrt{n}}
	\left(\imath_{X^n}(X^n)-nH\right)\geq
	Q^{-1}(\epsilon)+\frac{C}{\sigma\sqrt{n}}
	\right]\\
&\leq&
	Q\Big(Q^{-1}(\epsilon)
	+\frac{C}{\sigma\sqrt{n}}\Big)
	+\frac{A}{\sqrt{n}}\, ,
	\label{ikea}
\end{eqnarray}
where \eqref{ikea} follows from Theorem~\ref{thm:MarkovBE}.
Since,
\begin{eqnarray}
Q'(x)\!\!&=&\!\!-\phi(x) \\
0\leq Q''(x)\!\!&=&\!\!x\phi(x)\leq \frac{1}{\sqrt{2\pi e}}, ~~~x\geq 0,
\end{eqnarray}
a second-order Taylor expansion of  the first term 
in the right side of \eqref{ikea} gives,
\begin{eqnarray}
	\mathbb{P}[ \ell({\mathsf f}_n^*(X^n)) \geq K_n ]
&\leq&
	\epsilon
	-\frac{C}{\sigma\sqrt{n}}
	\phi(Q^{-1}(\epsilon))
	+
	\frac{1}{2\sqrt{2\pi e}}
	\Big(
	\frac{C}{\sigma\sqrt{n}}\Big)^2
	+\frac{A}{\sqrt{n}}
	\\
&\leq&
	\epsilon
	-\frac{1}{\sigma\sqrt{n}}
	\left\{
	C
	\Big[\phi(Q^{-1}(\epsilon))
	-
	\frac{1}{2\sqrt{2\pi e}}
	\Big(
	\frac{C}{\sigma\sqrt{n}}\Big)
	\Big]
	-A\sigma
	\right\},
	\label{eq:messM}
\end{eqnarray}
and choosing $C$ as in \eqref{thechosen} for $n$ satisfying \eqref{tribe}
the right side of (\ref{eq:messM})
is bounded above by $\epsilon$. 
Therefore, 
$\mathbb{P}[ \ell({\mathsf f}_n^*(X^n)) > K_n ]\leq\epsilon$,
which, by definition implies that $nR^*(n,\epsilon)\leq K_n$,
as claimed.
\end{IEEEproof}

\medskip

\begin{IEEEproof}[Proof of Theorem~\ref{thm:markov-converse}]
Applying Theorem~\ref{thm:conv1} with $X^n$ 
in place of $X$ 
and with $\delta>0$ and $K_n\geq 1$ arbitrary,
we obtain, 
\begin{eqnarray}
\mathbb{P}[\ell({\mathsf f}_n^*(X^n))\geq K_n]
&\geq&
	\mathbb{P}\Big[\imath_{X^n}(X^n)\geq K_n+\delta\Big]
	-2^{-\delta}\\
&=&
	\mathbb{P}\Big[ \frac{1}{\sigma\sqrt{n}}
	\left(\imath_{X^n}(X^n)-nH\right)\geq
	\frac{K_n-nH+\delta}{\sigma\sqrt{n}}
	\Big]-2^{-\delta}\\
&\geq&
	Q\Big(
	\frac{K_n-nH+\delta}{\sigma\sqrt{n}}
	\Big)
	-
	\frac{A}{\sqrt{n}}
	-2^{-\delta}, \label{miller}
\end{eqnarray}
where \eqref{miller} now follows from
Theorem~\ref{thm:MarkovBE}. 
Letting $\delta=\delta_n=\frac{1}{2}\log_2 n$ and,
\begin{eqnarray}K_n=
nH+\sigma\sqrt{n} Q^{-1}(\epsilon )
-\delta
-\frac{\sigma(A+1)}{\phi(Q^{-1}(\epsilon))},\end{eqnarray}
yields,
\begin{eqnarray}
\mathbb{P}[\ell({\mathsf f}_n^*(X^n))\geq K_n]
\geq
Q\Big(Q^{-1}(\epsilon)
-\frac{(A+1)}{\phi(Q^{-1}(\epsilon))\sqrt{n}}
\Big)-\frac{1}{\sqrt{n}}.\end{eqnarray}
Note that, since $\epsilon\in(0,1/2)$, we
have $Q^{-1}(\epsilon)>0$. And since
$Q'(x)=-\phi(x)$, a simple
two-term Taylor expansion of $Q$ above gives,
\begin{eqnarray}
\mathbb{P}[\ell({\mathsf f}_n^*(X^n))\geq K_n]
\geq
\epsilon
+\frac{A}{\sqrt{n}}\;>\;\epsilon,
\end{eqnarray}
for all,
\[
n\geq\left(
\frac{A+1}{Q^{-1}(\epsilon)\phi(Q^{-1}(\epsilon))}
\right)^2,
\] 
hence
$nR^*(n,\epsilon)> K_n-1$,
as claimed.
\end{IEEEproof}

\newpage

\section{Source Dispersion and Varentropy} \label{s:dispersion}

Traditionally, refined analyses in lossless data compression 
have focused attention
on the {\it redundancy},
defined as the difference between the minimum average 
compression rate and the entropy rate.
As we mentioned in Section \ref{sec:ear}, if the source statistics are known, 
then the per-symbol
redundancy is positive and behaves as $O\left(\frac{1}{n}\right)$
when the prefix condition is enforced, while it is 
 $- \frac{1}{2n} \log_2 n +  O (\frac{1}{n})$, without the prefix condition.
But since, as we saw
in Sections~\ref{section:memoryless} 
and~\ref{section:markov},
the standard deviation of the 
best achievable compression
rate is  $O (\frac{1}{\sqrt{n}})$,
the rate will be dominated by these
fluctuations. Therefore, as noted
in \cite{kontoyiannis-97}, it of primary 
importance to analyze the variance of 
the optimal codelengths.
To that end, we introduce the 
following operational definition:

\medskip

\begin{definition}
The dispersion $D$ (measured in bits$^2$) 
of a source $\{ P_{X^n} \}_{n=1}^{\infty}$ is,
\begin{eqnarray}
D = \limsup_{\ngi} \frac{1}{n} 
\VAR ( \ell ( \mathsf{f}^*_n (X^n ) )),
\label{eq:dispersion}
\end{eqnarray}
where $\ell ( \mathsf{f}^*_n ( \cdot ) )$ is the length of 
the optimum fixed-to-variable lossless code (cf.\ Section \ref{sec:ofv}).
\end{definition}

\medskip

As we show in Theorem~\ref{thm:dispersion} below,
for a broad class of sources, the dispersion $D$
is equal to the source varentropy rate $\sigma^2$
defined in (\ref{eq:ve}). Moreover, in view
of the Gaussian approximation bounds for $R^*(n,\epsilon)$
in Sections~\ref{section:memoryless} 
and~\ref{section:markov} -- and more generally,
as long as a similar two-term Gaussian
approximation in terms of the entropy rate
and varentropy rate can be established up to
$o(1/\sqrt{n})$ accuracy -- we can conclude
the following: by the definition of 
$n^*(R,\epsilon)$ in Section~\ref{sec:ofv},
the source blocklength $n$ required for the compression 
rate  
to exceed $(1 + \eta ) H$ with probability
no greater than $\epsilon > 0$ is approximated by,
\begin{align}
n^* ( (1 + \eta ) H, \epsilon ) &\approx 
\frac{\sigma^2}{H^2} \left( \frac{Q^{-1} (\epsilon)}{ 1 + \eta } \right)^2
\\
\label{cagando}
 &=
\frac{D}{H^2} \left( \frac{Q^{-1} (\epsilon)}{ 1 + \eta } \right)^2,
\end{align}
i.e., by the product of a factor that depends only on the source 
(through $H$ and $D$ or $\sigma^2$),
and a factor that depends only on the 
design requirements $\epsilon$ and $\eta$.
Note that this is in close parallel with the notion 
of channel dispersion introduced in \cite{PPV}.

\medskip

\begin{example}
Coin flips with bias $p$ have varentropy,
\begin{equation}
\sigma^2 = p (1-p) \log^2 \frac{1-p}{p},
\end{equation}
so 
the key parameter in \eqref{cagando} which characterizes 
the time horizon required for
the source to become ``typical" is,
\begin{eqnarray}
\frac{D}{H^2} = \frac{p - p^2}{\left(p + 
\frac{1}{\frac{\log p}{\log (1-p)} - 1}\right)^2}.
\end{eqnarray}
\end{example}

\medskip

\begin{example}
For a memoryless source whose marginal 
is the geometric distribution,
\begin{eqnarray}
P_{X} (k ) = q ( 1 - q)^k,
\end{eqnarray}
the ratio of varentropy to squared entropy is,
\begin{eqnarray}
\frac{\sigma^2}{H^2} 
=\frac{\sigma^2}{H^2} 
= (1-q) \left( \frac{\log_2( 1-q) }{h(q)}\right)^2,
\end{eqnarray}
where $h$ denotes the binary entropy function.
\end{example}

\begin{figure}[ht]
\centerline{\includegraphics[width=14cm]{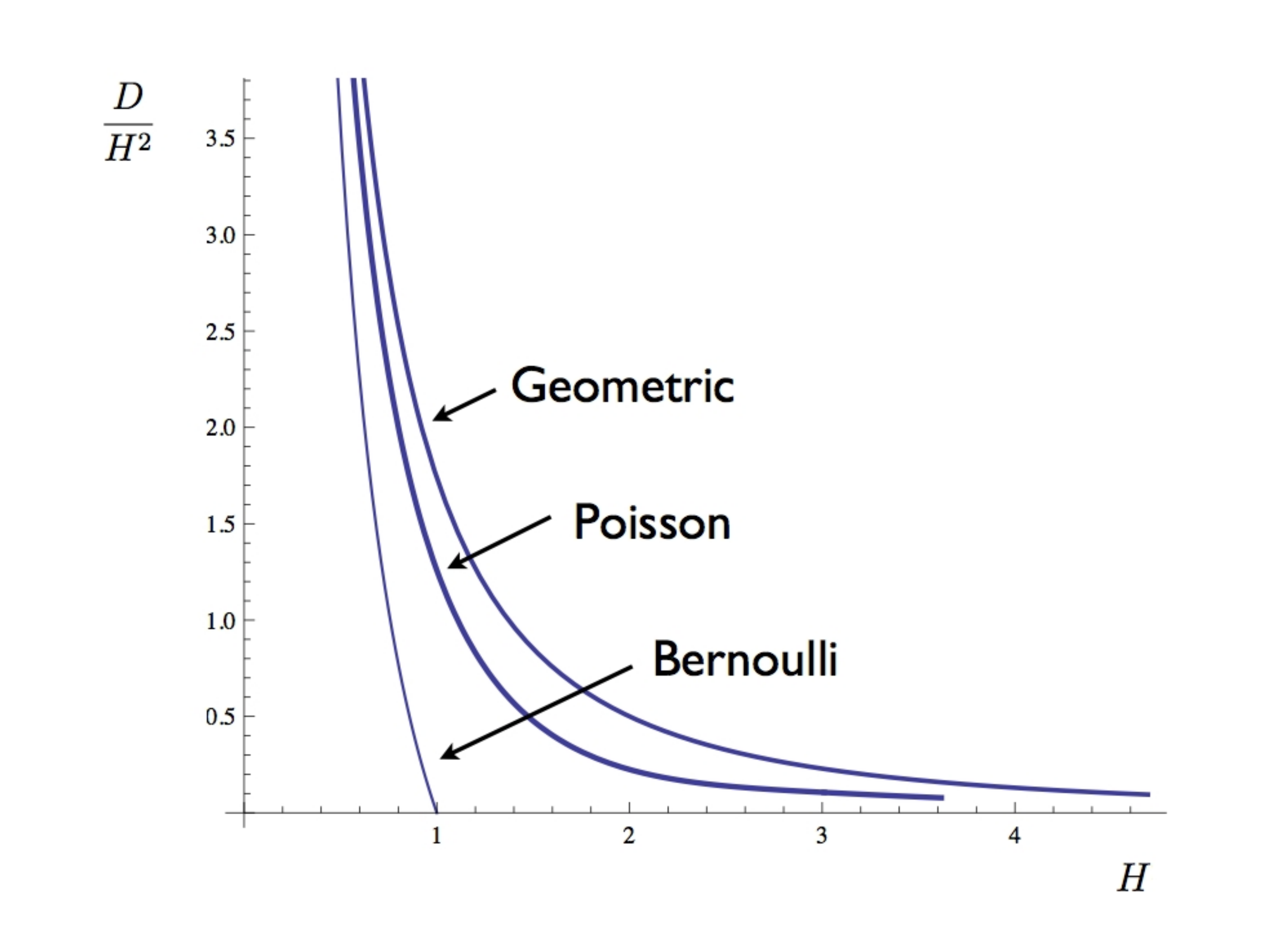}}
\caption{\label{fig:nordis} Normalized dispersion as a function of entropy for memoryless sources}
\end{figure}

Figure \ref{fig:nordis} compares the normalized dispersion to the entropy for the Bernoulli, geometric and Poisson
distributions. We see that as the source becomes more compressible (lower entropy per letter), the longer the horizon
over which we need to compress in order 
to squeeze most of the redundancy out of the source.

\medskip

\begin{definition}
A source $\{X_n\}$ 
taking values on the finite alphabet ${\cal A}$
is a {\em linear information growth} source
if any nonzero-probability string
has probability bounded below by
an exponential, that is, if 
there is a finite constant
$A$ and and an integer $N_0\geq 1$ such that,
for all $n\geq N_0$, every nonzero-probability string
$x^n\in{\cal A}^n$ satisfies
\begin{equation}
\imath_{X^n}(x^n)\leq An .
\label{eq:f-e}
\end{equation}
\end{definition}

\medskip

Any memoryless source belongs to the class of linear information growth.
Also note that, every irreducible and 
aperiodic Markov chain is a linear information growth
source: Writing $q$ for the smallest nonzero
element of the transition matrix, and $\pi$
for the smallest nonzero probability 
for $X_1$, we easily see that (\ref{eq:f-e})
is satisfied with $N_0=1$, 
$A=\log_2(1/q\pi)$.
The class of linear information growth sources
is related, at least at the level of
intuition, to the class of finite-energy
processes considered by Shields \cite{shields:book}
and to processes satisfying the
Doeblin-like condition of Kontoyiannis
and Suhov \cite{kontoyiannis-suhov}. 

We proceed to show an interesting regularity result for linear information growth sources:

\medskip

\begin{lemma}
\label{l:f-e}
Suppose $\{X_n\}$ is a 
(not necessarily stationary or ergodic)
linear information growth
source.
Then:
\begin{equation} \label{beam}
\lim_{n\to\infty}\frac{1}{n}{\mathbb E}
\Big[\Big(\ell({\mathsf f}_n^*(X^n))-\imath_{X^n}(X^n)\Big)^2\Big]=0.
\end{equation}
\end{lemma}

\medskip

\begin{IEEEproof}
For brevity, denote
$\ell_n = \ell(\mathsf{f}^*_n(X^n))$ and
$\imath_n = \imath_{X^n}(X^n)$,
respectively.
Select an arbitrary $\tau_n$. The expectation of interest is
\begin{equation}
{\mathbb E}[(\ell_n-\imath_n)^2]
=
	{\mathbb E}[(\ell_n-\imath_n)^2
	{\mathbb I}\{\ell_n\geq \imath_n-\tau_n\}]
	+
	{\mathbb E}[(\ell_n-\imath_n)^2
	{\mathbb I}\{\ell_n<\imath_n-\tau_n\}].
\end{equation}
Since
$\ell_n\leq \imath_n$,  on the event
$\{\ell_n\geq \imath_n-\tau_n\}$,
we have 
$(\ell_n-\imath_n)^2\leq\tau_n^2$.
Also, 
by the linear information growth assumption
we have the bound
$0\leq \imath_n-\ell_n\leq \imath_n\leq Cn$ for
a finite constant $C$ and all $n$
large enough. Combining these two
observations with Theorem~\ref{thm:conv2},
we obtain that,
\begin{eqnarray}	
{\mathbb E}[(\ell_n-\imath_n)^2]
&\leq&
	\tau_n^2
	+
	C^2n^2
	{\mathbb P}\{\ell_n<\imath_n-\tau_n\}\\
&\leq&
	\tau_n^2
	+
	C^2n^2 2^{-\tau_n}
	\left( n\log_2|{\cal A}|+1\right)\\
&\leq&
	\tau_n^2
	+
	C'n^3 2^{-\tau_n},
	\label{eq:lemma-step}
\end{eqnarray}
for some $C'<\infty$ and all $n$ large enough.
Taking $\tau_n=3\log_2 n$, dividing by $n$
and letting $n\to\infty$ gives the claimed
result.
\end{IEEEproof}

\medskip

Note that we have actually proved a stronger result, namely,
\begin{equation}\label{true}
\mathbb{E} \Big[\Big(\ell({\mathsf f}_n^*(X^n))-\imath_{X^n}(X^n)\Big)^2\Big] = O ( \log^2 n ).
\end{equation}
Linear information growth is sufficient for dispersion to equal varentropy:

\medskip

\begin{theorem}
\label{thm:dispersionLIG}
If the source has linear information growth, and finite varentropy, 
then:
\begin{eqnarray}
D = \sigma^2.
\end{eqnarray}
\end{theorem}

\medskip

\begin{IEEEproof}
For notational convenience, we abbreviate $H_n$ for $H(X^n)$.
Expanding the definition of the variance
of $\ell_n$, we obtain,
\begin{align}
&
	\VAR(\ell( \mathsf{f}^*_n (X^n ) ))	
	\nonumber\\
&=
	{\mathbb E}
	\left[
	\left(
	  \ell_n- {\mathbb E}[\ell_n]
	\right)^2
	\right]\\
&=
	{\mathbb E}
	\Big[
	\Big(
	(\ell_n-\imath_n)
	+(\imath_n-H_n)
	- \mathbb{E} [ \ell_n -\imath_n] )
	\Big)^2
	\Big]\\
&=
	\mathbb{E}[(\ell_n-\imath_n)^2]
	+{\mathbb E}[(\imath_n-H_n)^2]
	- \mathbb{E}^2 [ \ell_n -\imath_n] 
	+2{\mathbb E}[(\ell_n-\imath_n)(\imath_n-H_n)]
	\end{align}
and therefore, using the Cauchy-Schwarz inequality twice,
\begin{align}
&	|\VAR(\ell( \mathsf{f}^*_n (X^n ) ))
	-\VAR(\imath_{X^n}(X^n))|
	\nonumber\\
& =  | \mathbb{E}[(\ell_n-\imath_n)^2] - \mathbb{E}^2 [ \ell_n -\imath_n] + 2{\mathbb E}[(\ell_n-\imath_n)(\imath_n-H_n)]  | \\
&	\leq
	2{\mathbb E}[(\ell_n-\imath_n)^2]
	+2\left\{
	{\mathbb E}[(\ell_n-\imath_n)^2]
	\right\}^{1/2}
	[\VAR(\imath_{X^n}(X^n))]^{1/2}.
	\label{eq:varexpand}
\end{align}
Dividing by $n$ and letting $n\to\infty$, 
we obtain that the first term
tends to zero by Lemma~\ref{l:f-e}, 
and the second term becomes,
	\begin{equation}
	2
	\left\{
	\frac{{\mathbb E}[(\ell_n-\imath_n)^2]}{n}
	\right\}^{1/2}
	\left\{
	\frac{\VAR(\imath_{X^n}(X^n))}{n}
	\right\}^{1/2},
	\label{eq:disp-step}
	\end{equation}
which also tends to zero by Lemma~\ref{l:f-e} and
the finite-varentropy rate assumption. Therefore,
\begin{equation}
\lim_{n\to\infty}
\frac{1}{n}
	|\VAR(\ell( \mathsf{f}^*_n (X^n ) ))
	-\VAR(\imath_{X^n}(X^n))| =0,
\end{equation}
which, in particular, implies that $\sigma^2=D$.
\end{IEEEproof}

\medskip

In view of \eqref{true}, if  we normalize by $\sqrt{n} \log n$,
instead of $n$ in the last step of the proof of Theorem \ref{thm:dispersionLIG}, we obtain the stronger result:
\begin{equation}
\label{eq:theoremS}
|\VAR ( \ell ( \mathsf{f}^*_n (X^n ) ))-
\VAR ( \imath_{X^n} (X^n ) ))|
=O\left(\sqrt{n}\log_2 n\right).
\end{equation}
Also,
Lemma~\ref{l:f-e} and Theorem \ref{thm:dispersionLIG} remain valid if
instead of the linear information growth condition
we invoke the weaker assumption that there
exists a sequence
$\tau_n = o ( \sqrt{n} ) $, 
such that,
\begin{equation}
\max_{x^n \colon P_{X^n}(x^n)\neq 0} \imath_{X^n}(x^n)
=
o\left(2^{\epsilon_n / 2}\right).
\label{eq:weakC}
\end{equation}

We turn now attention to the Markov chain case.

\medskip

\begin{theorem}
\label{thm:dispersion}
Let $\{X_n\}$ be an irreducible, aperiodic
(not necessarily stationary) Markov source
with entropy rate $H$. Then:
\begin{enumerate}
\item
The varentropy rate $\sigma^2$ defined in (\ref{eq:ve})
exists as the limit,
\begin{equation}
\sigma^2 = \lim_{n\to\infty} \frac{1}{n} 
\VAR ( \imath_{X^n} (X^n ) )).
\end{equation}
\item
The dispersion $D$ defined in (\ref{eq:dispersion})
exists as the limit,
\begin{eqnarray}
D = \lim_{\ngi} \frac{1}{n} 
\VAR ( \ell ( \mathsf{f}^*_n (X^n ) )).
\label{eq:Dlim}
\end{eqnarray}
\item
$D=\sigma^2$.
\item
The varentropy rate (or, equivalently, the dispersion)
can be characterized in terms of the best
achievable rate $R^*(n,\epsilon)$ as,
\begin{eqnarray}
\sigma^2 
= \lim_{\epsilon \rightarrow 0} \lim_{\ngi} 
\frac{n \left( R^*(n, \epsilon) - H \right)^2}{2 \ln \frac{1}{\epsilon}}
= \lim_{\epsilon \rightarrow 0} \lim_{\ngi} 
n\left(\frac{R^*(n, \epsilon) - H}{Q^{-1}(\epsilon)}\right)^2,
\label{eq:RD}
\end{eqnarray}
as long as $\sigma^2$ is nonzero.
\end{enumerate}
\end{theorem}

\medskip

\begin{IEEEproof} 
The limiting expression in part~$1)$ was already
established in  Theorem~\ref{thm:CLT} of 
Section~\ref{section:asymptotic};
see also the discussion leading to~(\ref{eq:vel})
in Section~\ref{section:markov}. 
Recalling that every irreducible
and aperiodic Markov source is a linear information growth
source, combining part~$1)$
with Theorem \ref{thm:dispersionLIG} immediately yields the results of 
parts~$2)$ and $3)$.

Finally, part~$4)$ follows from the results
of Section~\ref{section:markov}. Under the 
present assumptions, 
Theorems~\ref{thm:markov-direct} 
and~\ref{thm:markov-converse} together
imply that there is a finite constant $C_1$ 
such that,
\begin{equation}
\Big|\sqrt{n}(R^*(n,\epsilon)-H) -
\sigma Q^{-1}(\epsilon)\Big|\leq \frac{1}{2}\frac{\log_2 n}{\sqrt{n}}
+\frac{C_1}{\sqrt{n}},
\end{equation}
for all $\epsilon\in(0,1/2)$ and all $n$ large enough.
Therefore,
\begin{equation}
\lim_{n\to\infty}
n(R^*(n,\epsilon)-H)^2=
\sigma^2 (Q^{-1}(\epsilon))^2.
\end{equation}
Dividing by $2\ln\frac{1}{\epsilon}$,
letting $\epsilon\downarrow 0$,
and recalling the simple fact that
$(Q^{-1}(\epsilon))^2\sim 2\ln\frac{1}{\epsilon}$
(see, e.g., \cite[Section~3.3]{verdu-book})
proves~(\ref{eq:RD}) and completes the proof of the
theorem.
\end{IEEEproof}

\medskip

From Theorem~\ref{thm:dispersionLIG} it 
follows that, for a broad class of sources
including all ergodic Markov chains with nonzero 
varentropy rate,
\begin{eqnarray} 
\label{ratiovariances}
\lim_{n \to \infty}
\frac{\VAR \left(  \ell ( {\mathsf  f }_n^* (X^n  )) \right) }
{\VAR \left( \imath_{X^{n}} ( X^{n}) \right)}
= 1.
\end{eqnarray}
Analogously to Theorem~\ref{thm:meanratio}, 
we could explore 
whether~\eqref{ratiovariances} might hold 
under broader conditions, including
the general setting of possibly non-serial 
sources. However, consider the 
following simple example.

\medskip

\begin{example}\label{example:equiprobable:var}
As in Example \ref{example:equiprobable}, let $X_M$ 
be equiprobable on a set of $M$ elements, then,
\begin{eqnarray}
H (X_M) &=& \log_2 M \\
\VAR \left( \imath_{X_M} ( X_M) \right) &=& 0 \\
\limsup_{M \rightarrow \infty} \VAR \left( \ell ( {\mathsf  f }^* ( X_M  )) \right) &=& 2 + \frac14  \label{gardenSUP}\\
\liminf_{M \rightarrow \infty} \VAR \left( \ell ( {\mathsf  f }^* ( X_M  )) \right) &=& 2.  \label{gardenINF}
\end{eqnarray}
To verify \eqref{gardenSUP} and \eqref{gardenINF},
define the function,
\begin{eqnarray}
s (K) &=& \sum_{i=1}^K i^2 \, 2^i \\
&=& - 6 +  2^{K+1} ( 3 - 2 K  + K^2 ).
\end{eqnarray}
It is straightforward to check that,
\begin{eqnarray}
\mathbb{E} \left[ \ell^2 ( {\mathsf  f }^* ( X_M  )) \right]
=
\frac{1}{M} 
\left( 
s ( \lfloor \log_2 M \rfloor )
-
( \lfloor \log_2 M \rfloor )^2
\cdot \left( 2^{\lfloor \log_2 M \rfloor + 1} - M -1 \right)
\right). \label{theseus}
\end{eqnarray}
Together with \eqref{humongous}, 
\eqref{theseus}
results in,
\begin{eqnarray}
\VAR \left( \ell ( 
{\mathsf  f }^* ( X_M  )) \right) = 3 \xi_M - \xi_M^2 + o (1),
\end{eqnarray}
with,
\begin{eqnarray}
\xi_M = \frac{2^{1+\lfloor \log_2 M \rfloor}}{M},
\end{eqnarray}
which takes values in $(1,2]$. On that 
interval, the parabola $3 x - x^2$ takes a minimum value of 2
and a maximum value of $(3/2)^2$, 
and \eqref{gardenSUP}, \eqref{gardenINF} follow.
\end{example}

\medskip

Although the ratio of optimal codelength variance to 
the varentropy rate may be infinity
as illustrated in Example \ref{example:equiprobable:var},
we do have the following counterpart of the first-moment 
result in Theorem \ref{thm:meanratio}
for the second moments:

\medskip

\begin{theorem}
\label{thm:varratio}
 For any (not necessarily serial) source $\mathbf{X} = \{P_{X^{(n)}}\}_{n=1}^\infty$,
\begin{eqnarray} \label{remindme}
\lim_{n \to \infty}
\frac{ \mathbb{E}  [ \ell^2 ( {\mathsf  f }_n^* (X^{(n)}  ))  ]}{\mathbb{E} \left[ \imath^2_{X^{(n)}} ( X^{(n)})  \right]} = 1,
\end{eqnarray}
as long as the denominator diverges.
\end{theorem}

\medskip

\begin{IEEEproof}
Theorem \ref{thm:simple} implies that,
\begin{eqnarray}
\mathbb{E}  [ \ell^2 ( {\mathsf  f }_n^* (X^{(n)}  ))  ] \leq \mathbb{E} \left[ \imath^2_{X^{(n)}} ( X^{(n)})  \right].
\end{eqnarray}
Therefore, the $\limsup$ in 
\eqref{remindme} is bounded above by 1. 
To establish the corresponding lower bound,
fix an arbitrary
$\vartheta>0$. Then,
\begin{eqnarray}
\mathbb{E} [ \ell^2 ( {\mathsf  f }_n^* (X^{(n)}  )) ] 
& = & 
\sum_{k\geq 1}\mathbb{P}\left[\ell^2 ( {\mathsf  f }_n^* (X^{(n)}  )) \geq k\right]\\
& = & 
\sum_{k\geq 1}\mathbb{P}\left[\ell^{*}_n\geq \sqrt{k}\right]\\
& = & 
\sum_{k\geq 1}\mathbb{P}\left[\ell^{*}_n\geq \lceil\sqrt{k}\rceil\right]\\
& \geq & 
\sum_{k\geq 1}
\left[\mathbb{P}\left[\imath_{X^{(n)}} ( X^{(n)}) \geq (1+\vartheta)\lceil\sqrt{k}\rceil\right]
-2^{-\vartheta\lceil\sqrt{k}\rceil}\right], \label{koleh}
\end{eqnarray}
where \eqref{koleh} follows by letting $\tau = \vartheta\lceil\sqrt{k}\rceil$
in the converse Theorem~\ref{thm:conv1}. Therefore,
\begin{eqnarray}
\mathbb{E} [ \ell^2 ( {\mathsf  f }_n^* (X^{(n)}  )) ] 
& \geq & -C_\vartheta +
\sum_{k\geq 1}
\mathbb{P}\left[\imath^2_{X^{(n)}} ( X^{(n)})\geq (1+\vartheta)^2\lceil\sqrt{k}\rceil^2\right]
\label{pori}\\
& \geq & 
-D_\vartheta +
\sum_{k\geq 1}
\mathbb{P}\left[\frac{\imath^2_{X^{(n)}} ( X^{(n)})}{(1+\vartheta)^3}\geq k\right]
\label{fori}\\
&\geq&
-D_\vartheta +
\frac{1}{(1+\vartheta)^3}
\mathbb{E} [ \imath^2_{X^{(n)}} ( X^{(n)}) ].
\label{gori}
\end{eqnarray}
where $C_\vartheta$, $D_\vartheta$ are positive scalars that only vary with $\vartheta$.
Note that \eqref{pori} holds because $a^{-\sqrt{k}}$ is 
summable
for all $0 < a < 1$;
\eqref{fori} holds because $ (1 + \vartheta ) k \geq \lceil\sqrt{k}\rceil^2 $ for all sufficiently large $k$;
and \eqref{gori} holds because,
\begin{eqnarray}
\int_{k}^{k+1} (1 - F (x) ) \, dx \geq 1 - F(k+1),
\end{eqnarray}
whenever $F(x)$ is a cumulative distribution function.
Dividing both sides of \eqref{pori}-\eqref{gori} by 
the second moment $\mathbb{E} [ \imath^2_{X^{(n)}} ( X^{(n)}) ]$
and letting $n \to \infty$, we conclude that the ratio 
in \eqref{remindme} is lower bounded by $(1 + \vartheta)^{-3}$.
Since $\vartheta$ can be taken to 
be arbitrarily small, this proves that the 
$\liminf$ \eqref{remindme} is lower bounded by 1,
as required.
\end{IEEEproof}


\bigskip

\begin{center}
\sc{Acknowledgments}
\end{center}
The work of SV was supported 
in part by the National Science Foundation (NSF) under Grant 
CCF-1016625 and by the Center for Science of Information (CSoI), 
an NSF Science and Technology Center, under Grant CCF-0939370.
The work of IK was supported in part by 
the research program `CROWN' through the
Operational Program `Education and Lifelong Learning 2007-2013' of NSRF.
Parts of this paper were presented at the 46th Annual Conference on 
Information Sciences and Systems, Princeton University, Princeton, NJ, 
March 21-23, 2012.


\end{document}